\documentclass[a4paper,11pt]{article}
\usepackage[dvipdfmx]{color}
\usepackage[dvipdfmx]{graphicx}
\usepackage{mathrsfs,braket,amsmath,amssymb}
\usepackage{jheppub}
\usepackage[T1]{fontenc} 

\def\nn{\nonumber}

\def\beq{\begin{equation}}
\def\eeq{\end{equation}}
\def\bea{\begin{eqnarray}}
\def\eea{\end{eqnarray}}
\def\bsub{\begin{subequations}}
\def\esub{\end{subequations}}

\def\C{{\cal C}}
\def\S{{\cal S}}
\def\nn{\nonumber}
\def\beq{\begin{equation}}
\def\eeq{\end{equation}}
\def\bei{\begin{itemize}}
\def\eei{\end{itemize}}
\def\bea{\begin{eqnarray}}
\def\eea{\end{eqnarray}}

\def\SKL{S_{\psi K_L}}
\def\SKS{S_{\psi K_S}}
\def\GKL{G_{\psi K_L}}
\def\GKS{G_{\psi K_S}}
\def\CKL{C_{\psi K_L}}
\def\CKS{C_{\psi K_S}}
\title{\boldmath Precise Discussion 
of Time-Reversal Asymmetries in B-meson decays}

\author[]{Takuya Morozumi,}
\author[]{Hideaki Okane}
\author[]{and Hiroyuki Umeeda}

\affiliation[]{Graduate School of Science, 
Hiroshima University,\\
Higashi-Hiroshima,
739-8526, Japan
}

\emailAdd{morozumi@hiroshima-u.ac.jp}
\emailAdd{hideaki-ookane@hiroshima-u.ac.jp}
\emailAdd{umeeda@theo.phys.sci.hiroshima-u.ac.jp}

\abstract{
BaBar collaboration announced that they observed time reversal (T) asymmetry through $B$ meson system.
In the experiment, 
time dependencies of two distinctive processes,
$B_- \rightarrow \bar{B^0}$ and $\bar{B^0}\rightarrow B_-$
($-$ expresses CP value)
are compared with each other.
In our study,
we examine
event number difference of these two processes.
In contrast to the BaBar asymmetry, the asymmetry of events
number includes the overall normalization difference for rates.
Time dependence of the asymmetry
is more general and it
includes terms absent
in one used by BaBar collaboration.
Both of the BaBar asymmetry and ours
are naively thought to be T-odd since two processes compared are related with flipping time direction.
We investigate the
time reversal transformation property
of our asymmetry.
Using our notation, one can see that
the asymmetry is not precisely a T-odd quantity,
taking into account indirect CP and CPT violation of K meson systems.
The effect of $\epsilon_K$ is extracted and gives rise to
$\mathcal{O}(10^{-3})$ contribution.
The introduced parameters are invariant under rephasing
of quarks
so that the coefficients of our asymmetry
are expressed as phase convention independent quantities.
Some combinations of the asymmetry enable us
to extract parameters for wrong sign decays of $B_d$ meson,
CPT violation, etc.
We also study the reason why the T-even terms
are allowed to contribute to the asymmetry,
and find that
several conditions are needed for
the asymmetry to be a T-odd quantity.}
\keywords{B-Physics, CP violation, Kaon Physics}
\arxivnumber{1411.2104}
\begin{document}
\preprint{HUPD1404}
\maketitle
\flushbottom
\section{Introduction}
\label{intro}
T-symmetry is fundamental symmetry in particle physics.
T-transformation exchanges an initial state and final state, flipping the momentum and spin of particles.
If some rate of a process deviates from the rate of time reversed one,
it implies T-violation.
T-violation is worth pursuit since it reflects the characterized feature of theory.
\par
In EPR correlating B meson system, if one of a pair of B meson is tagged,
another side of B meson is determined as orthogonal state
with tagged side B meson.
In refs.\cite{Banuls:1999aj}-\cite{Bernabeu:2012ab},
a method to observe T-violation using B meson system is suggested.
Their idea is based on
the difference of the time dependencies for
two distinctive processes,
$B_{-}\rightarrow \bar{B^0}$ and $\bar{B^0}\rightarrow B_{-}$.
In refs.\cite{Banuls:1999aj}-\cite{Bernabeu:2012ab},
it
is considered to be T-asymmetry since two distinctive processes are related with flipping time direction.
Then, BaBar collaboration announced \cite{Lees:2012uka}
that they measured
non-zero asymmetry
and this observation is direct demonstration of T-violation.
A review for the BaBar asymmetry and the other T and CPT asymmetries is given in ref. \cite{Schubert:2014ska}.
\par
However, their statement includes ambiguity since we cannot
exactly identify B meson such as $\bar{B^0}$ or $B_-$.
In the BaBar experiment, two methods to identify $\bar{B^0}$ and $B_-$ are implemented.
The first one is referred as flavor tagging which enables one
to identify $\bar{B^0}$, using the semi-leptonic decay mode of B meson.
Another one is referred as CP tagging which allows us to
identify $B_{-}$ with final state $\psi K_S$.
\par
In ref.\cite{Applebaum:2013wxa}, it is pointed out that
there exist subtleties in BaBar measurement.
The main point in ref.\cite{Applebaum:2013wxa} is that
they consider a process and its authentically time reversed process.
Note that the authentic time reversed process includes
inverse decay such like $l^-X\rightarrow \bar{B^0}$
and $\psi K_L\rightarrow B^0$.
Since an authentic time reversed process is experimentally
hard to observe,
they substitute another process which does not include
inverse decays.
Therefore, they derive the conditions that
BaBar asymmetry is identical with a T-odd quantity,
taking into account inverse decays.
The derived conditions are (1) the absence of the wrong sign
semi-leptonic B meson decays,
(2) the absence of the wrong strangeness decays,
(3) the absence of CPT violation of the strangeness changing decays.
All the conditions are derived by assuming that
$\psi K_S$ and $\psi K_L$ are exact CP eigenstates.
\par
In this paper, we conduct model-independent analysis of
an event number asymmetry. Our analysis is extension of the work \cite{Applebaum:2013wxa},
incorporating the difference of overall constants for the rates
that form the asymmetry into calculation.
We also include the effect of indirect CP and CPT violation of K meson system.
Furthermore, the asymmetry is written in terms of the phase convention independent
parameters and one can find contribution of $\epsilon_K$ explicitly.
Some combinations of the coefficients
enable one to constrain model-independent parameters.
We also discuss the T-even parts of the asymmetry.
One can find that the asymmetry is a T-odd quantity
when several conditions are satisfied.
\par
In section \ref{Sec:1}, we introduce the asymmetry of entangled B meson system.
In section \ref{Sec:3},
we define parameters which are definitively T-odd or T-even
and describe the relation between notation in 
\cite{Applebaum:2013wxa} and ours.
It is also shown that the parameters are phase convention independent quantities.
The method to extract the effect of indirect CP violation
in Kaon system is also considered in section \ref{Sec:3}.
In section \ref{Sec:2}, we write the
event number
asymmetry in terms of the parameters defined in section \ref{Sec:3}
and show that the asymmetry consists of
not only the T-odd part but also T-even part,
using our notation.
In sections \ref{sub41}-\ref{sub43},
rather than discussing T-transformation property of the asymmetry,
some methods to extract the parameters from
the asymmetry are investigated.
In section \ref{Sec:4}, we derive the conditions that T-even 
parts of the asymmetry vanish and
examine the intuitive reason why these conditions are 
required.
The conditions are categorized as two types.
The first condition is in regards to
B meson states that appear in the processes.
Including the effect of indirect CP violation,
we evaluate how the first condition is violated,
in comparison with the result in \cite{Applebaum:2013wxa}.
The second condition is in regards to overall constant which
forms the asymmetry.
We find that the second condition is needed when
one takes account of the difference of overall constant of
the two rates.
Section \ref{Conclusion} is devoted to summary of our study.
\section{Formula for asymmetry of entangled systems}
\label{Sec:1}
In ref.\cite{Applebaum:2013wxa}, a formula for the time-dependent decay rate of the entangled $B\bar B$ system is derived.  When $f_1$ and $f_2$ denote the final states of a tagging side and a signal
side, respectively, it is written as, 
\bea
\Gamma_{(f_1)\perp, f_2}=e^{-\Gamma(t_1+t_2)}N_{(1)\perp, 2}
[\kappa_{(1)\perp, 2}\cosh(y\Gamma t)+\sigma_{(1)\perp, 2}\sinh(y\Gamma t)
\nn\\
+\C_{(1)\perp, 2}\cos(x\Gamma t)+\S_{(1)\perp, 2}\sin(x\Gamma t)]\nn\\
=e^{-\Gamma(t_1+t_2)}N_{(1)\perp, 2}\kappa_{(1)\perp, 2}
[\cosh(y\Gamma t)+\frac{\sigma_{(1)\perp, 2}}{\kappa_{(1)\perp, 2}}\sinh(y\Gamma t)
\nn\\
+\frac{\C_{(1)\perp, 2}}{\kappa_{(1)\perp, 2}}\cos(x\Gamma t)+\frac{\S_{(1)\perp, 2}}{\kappa_{(1)\perp, 2}}\sin(x\Gamma t)]
,\label{T1}
\eea
where
\bea
\Gamma=\frac{\Gamma_H+\Gamma_L}{2}
, \quad x=\frac{m_H-m_L}{\Gamma}
, \quad y=\frac{\Gamma_H-\Gamma_L}{2\Gamma}.
\label{T1p}
\eea
The expressions for $N_{(1)\perp, 2}, \kappa_{(1)\perp, 2}, \sigma_{(1)\perp, 2}, \C_{(1)\perp, 2}$ and $\S_{(1)\perp, 2}$ are given in ref.\cite{Applebaum:2013wxa}. 
For the sake of completeness, we record their expressions in eqs. (\ref{M1}-\ref{M5}).
Hereafter, we evaluate an asymmetry
including overall factor $N_{(1)\perp, 2}\kappa_{(1)\perp,2}$
in eq. (\ref{T1}).
One obtains a generic formula for the event number asymmetry of the two distinctive sets for final states;
$(f_1, f_2)$ versus $(f_3, f_4)$ as, 
\bea
A&\equiv& \frac{\Gamma_{(f_1)\perp, f_2}-\Gamma_{(f_3)\perp, f_4}}
{\Gamma_{(f_1)\perp, f_2}+\Gamma_{(f_3)\perp, f_4}}\nn\\
&=&\frac{(\frac{1}{\sqrt{N_R}}-\sqrt{N_R})\cosh(y\Gamma t)+\Delta\sigma \sinh(y\Gamma t)+\Delta \S
\sin(x\Gamma t)+\Delta \C\cos(x\Gamma t)}
{(\frac{1}{\sqrt{N_R}}+\sqrt{N_R})\cosh(y\Gamma t)+\hat{\sigma} \sinh(y\Gamma t)+\hat{\S}
\sin(x\Gamma t)+\hat{\C}\cos(x\Gamma t)},
\label{T2}
\eea
where
\bea
N_R&\equiv&\frac{N_{(3)\perp, 4}\kappa_{(3)\perp, 4}}{N_{(1)\perp, 2}\kappa_{(1)\perp, 2}},\label{T3}\\
\Delta X&\equiv&\frac{1}{\sqrt{N_R}}
\frac{X_{(1)\perp, 2}}{\kappa_{(1)\perp, 2}}
-\sqrt{N_R}\frac{X_{(3)\perp, 4}}{\kappa_{(3)\perp, 4}}\quad\quad(\mathrm{for}\: X=\sigma, \C, \S),\label{T4}\\
\hat{X}&\equiv&\frac{1}{\sqrt{N_R}}
\frac{X_{(1)\perp, 2}}{\kappa_{(1)\perp, 2}}
+\sqrt{N_R}\frac{X_{(3)\perp, 4}}{\kappa_{(3)\perp, 4}}\quad\quad(\mathrm{for}\: X=\sigma, \C, \S).\label{T5}
\eea
In eqs. (\ref{T2}-\ref{T5}), contribution from
different overall factors in eq. (\ref{T1}) for two processes are taken into account.
Taking the limit $N_R\rightarrow1$
and $y\rightarrow0$ in eq. (\ref{T2}),
one finds an asymmetry whose overall normalization
is eliminated, used in \cite{Lees:2012uka}.
In eq. (\ref{T4}-\ref{T5}),
$\Delta \mathcal{S}$ $(\Delta \mathcal{C})$ is equal to
$\Delta S_T^+$
$(\Delta C_T^+)$ defined in \cite{Applebaum:2013wxa}
when one takes the limit $N_R\rightarrow 1$.
In practice, we only need to consider the time difference $t$ within the interval
$[0,\tau_B]$ where $\tau_B$ is the life time of B meson.
Therefore, the approximation $\sinh(y\Gamma t)\simeq y\Gamma t, \cosh(y\Gamma t)\simeq 1$ is valid since
$y\ll 1$ for neutral B meson system.
Thus, we expand $A$ with respect to $y\Gamma t$,
\bea
A&\simeq&
\frac{\frac{1}{\sqrt{N_R}}-\sqrt{N_R}+\Delta\sigma y\Gamma t+\Delta \S
\sin(x\Gamma t)+\Delta \C\cos(x\Gamma t)}
{\frac{1}{\sqrt{N_R}}+\sqrt{N_R}+\hat{\sigma} y\Gamma t+\hat{\S}
\sin(x\Gamma t)+\hat{\C}\cos(x\Gamma t)}\nn\\
&=&
\frac{-\displaystyle\frac{\Delta N_R}{2}+\displaystyle\frac{\Delta\sigma}{2} y\Gamma t+\displaystyle\frac{\Delta \S}{2}
\sin(x\Gamma t)+\displaystyle\frac{\Delta \C}{2}\cos(x\Gamma t)}
{1+\displaystyle\frac{\hat{\sigma}}{2} y\Gamma t+\displaystyle\frac{\hat{\S}}{2}
\sin(x\Gamma t)+\displaystyle\frac{\hat{\C}}{2}\cos(x\Gamma t)}
,\label{T6}
\eea
where we denote
\bea
N_R=1+\Delta N_R
\label{T7}.
\eea

\begin{table}
\begin{center}
\begin{tabular}{|c|c||c|c|}\hline
Notation of this paper & Notation of \cite{Applebaum:2013wxa} &  
Notatin of \cite{Applebaum:2013wxa}
& Notation of this paper \\ \hline
$\lambda_{\psi K_S}$ & $\frac{p_K}{q_K} \lambda_{\psi K_1}$ & 
$\lambda_{\psi K_1}$ &  $\lambda^\prime(1-\Delta \lambda_{wst})$ 
\\ \hline 
$\lambda_{\psi K_L}$ &$\frac{p_K}{q_K} \lambda_{\psi K_2}$  & 
$\lambda_{\psi K_2}$ &  $-\lambda^\prime (1+\Delta \lambda_{wst})$ 
\\ \hline 
$G_{\psi K_S}$ & $G_{\psi K_1} - 2 S_{\psi K_1} \epsilon^I_K$ & 
$ \hat{G}_{\psi K}=\frac{G_{\psi K_1}-G_{\psi K_2}}{2}$& 
$G^\prime$ \\ \hline
$S_{\psi K_S}$ & $S_{\psi K_1} + 2 G_{\psi K_1} \epsilon^I_K$ & 
$\hat{S}_{\psi K}=\frac{S_{\psi K_1}-S_{\psi K_2}}{2}$ & 
$S^\prime$ \\ \hline
$C_{\psi K_S}$ & $C_{\psi K_1} -2 \epsilon^R_K$ & 
$\hat{C}_{\psi K}=\frac{C_{\psi K_1}+C_{\psi K_2}}{2}$
& $C^\prime$ \\ \hline
$G_{\psi K_L}$ & $G_{\psi K_2} - 2 S_{\psi K_2} \epsilon^I_K$ & 
$ \Delta G_{\psi K}=\frac{G_{\psi K_1}+G_{\psi K_2}}{2}$
& $S^\prime \Delta \lambda^I_{wst}$ \\ \hline
$S_{\psi K_L}$ & $S_{\psi K_2} + 2 G_{\psi K_2} \epsilon^I_K$ &$ \Delta S_{\psi K}=\frac{S_{\psi K_1}+S_{\psi K_2}}{2}$ &
$-G^\prime \Delta \lambda^I_{wst}$
 \\ \hline
$C_{\psi K_L}$ & $C_{\psi K_2} -2 \epsilon^R_K$ & 
$ \Delta C_{\psi K}=\frac{C_{\psi K_1}-C_{\psi K_2}}{2}$
& $\Delta \lambda^R_{wst}$
 \\ \hline
$\theta_K$ & 
$\hat{\theta}_{\psi K}=\frac{\theta_{\psi K1}+\theta_{\psi K2}}{2}$ 
&
$\Delta{\theta}_{\psi K}=
\frac{\theta_{\psi K1}-\theta_{\psi K2}}{2}$ & 0 
\\ \hline
\end{tabular}
\end{center}
\caption{The correspondence of parameters in this paper and those of
\cite{Applebaum:2013wxa}. 
In this paper, $\psi K_1$ corresponds to $\psi K_S$
and $\psi K_2$ corresponds to $\psi K_L$ respectively in 
\cite{Applebaum:2013wxa}
where indirect CP violation parameter $\epsilon_K$ is neglected.
In this paper,
$\psi K_L$ and $\psi K_S$ include the effect of indirect CP and CPT
violation. The first column 
shows the quantities defined
for mass eigenstates ($K_L, K_S$).  
From the third row to the eighth row in the second column, 
the quantities 
in the first column are expanded up to the first order of 
$\epsilon_K$  and are written in terms of the 
quantities for CP eigenstates $K_1, K_2$.
In the third column and in the fourth column, we show
how $(\hat{G}_{\psi K}, \hat{S}_{\psi K}, \hat{C}_{\psi K})$ and 
$(\Delta G_{\psi K}, \Delta S_{\psi K}, \Delta C_{\psi K}) $ 
in \cite{Applebaum:2013wxa} are related to
$(G^\prime, S^\prime, C^\prime, \Delta \lambda_{wst})$
 defined in this paper. About CPT violation parameter of strangeness changing decay, one can show $\theta_{\psi K1}=\theta_{\psi K2}=\theta_K$. Therefore 
$\hat{\theta}_{\psi K}=\theta_K$ and $\Delta \theta_{\psi K}=0$.
}
\label{tab:1}
\end{table}
\section{Parameter definitions in terms of flavor based state}
\label{Sec:3}
In this section, we introduce parameters that reveal in
the event number asymmetry in eq.(\ref{T6})
which we consider. In the processes which form the asymmetry,
final states of B-decay are given as the same ones used for the BaBar experiment\cite{Lees:2012uka}.
%
Mixing-parameters, $p, q, z, p_K, q_K$ and $z_K$ are defined
in appendix \ref{Sec:D}.
Eqs.(\ref{poverq}-\ref{zet}) lead the transformation property of
mixing parameter as,
$p\overset{\mathrm{CP}\:
\mathrm{or}\:\mathrm{T}}{\underset{}{\leftrightarrows}}q,\quad$
$p\xrightarrow{\mathrm{CPT}}p,\quad$
$q\xrightarrow{\mathrm{CPT}}q.$
Similarly, we obtain the transformation property of $z$ as,
$z\xrightarrow{\mathrm{CP}}-z, z\xrightarrow{\mathrm{T}}+z,
z\xrightarrow{\mathrm{CPT}}-z$.
The transformation properties of $p_K, q_K$ and $z_K$
are the same as $p, q$ and $z$, respectively.\\
Following 
\cite{Applebaum:2013wxa}, we introduce B meson decay amplitudes and
inverse decay amplitudes,
\bea
A_f\equiv\bra{f}T\ket{B^0}, \quad \bar{A}_f\equiv \bra{f}T\ket{\bar{B^0}},
\quad
A_f^{\mathrm{ID}}\equiv \bra{B^0}T\ket{f^T}, \quad 
\bar{A}_f^{\mathrm{ID}}\equiv \bra{\bar{B^0}}T\ket{f^T}.
\label{K2}
\eea
where $f^T$ is the time reversed state of $f$, i.e.,
the state with flipped momenta and spins.
Note that $A_f$ $(\bar{A}_f)$ and $A_f^{\mathrm{ID}}$
$(\bar{A}_f^{\mathrm{ID}})$
are exchanged under T-transformation.
Throughout this paper, we introduce
the notation $G_f, S_f$ and $C_f$
which are written in terms of amplitude ratio $\lambda_f$.
\bea
&G_{f}=
\displaystyle\frac{2\mathrm{Re}
\lambda_{f}}{1+|\lambda_{f}|^2},\quad
S_{f}=
\displaystyle\frac{2\mathrm{Im}
\lambda_{f}}{1+|\lambda_{f}|^2},\quad
C_{f}=\displaystyle\frac{1-|\lambda_{f}|^2}
{1+|\lambda_{f}|^2},&\label{K1pppp}\\
&G_f^2+S_f^2+C_f^2=1.&\label{sum2}
\eea
Using notation (\ref{K2}), we can denote following parameters,
\bea
&\lambda_{\psi K_{S, L}}\equiv
\displaystyle\frac{q}{p}
\displaystyle\frac{\bar{A}_{\psi K_{S, L}}}{A_{\psi K_{S, L}}}
\sqrt{\displaystyle\frac{1+\theta_{\psi K_{S, L}}}
{1-\theta_{\psi K_{S, L}}}}
=\displaystyle\frac{q}{p}
\displaystyle\frac{A_{\psi K_{S, L}}^{\mathrm{ID}}}
{\bar{A}_{\psi K_{S, L}}^{\mathrm{ID}}}
\sqrt{\displaystyle\frac{1-\theta_{\psi K_{S, L}}}
{1+\theta_{\psi K_{S, L}}}}
,&\label{K1}\\
&\theta_{\psi K_{S, L}}=
\displaystyle\frac{A_{\psi K_{S, L}}A_{\psi K_{S, L}}^{\mathrm{ID}}
-\bar{A}_{\psi K_{S, L}}\bar{A}_{\psi K_{S, L}}^{\mathrm{ID}}}
{A_{\psi K_{S, L}}A_{\psi K_{S, L}}^{\mathrm{ID}}
+\bar{A}_{\psi K_{S, L}}\bar{A}_{\psi K_{S, L}}^{\mathrm{ID}}},
&
\eea
Note that $\psi K_L$ and $\psi K_S$ are not exact CP eigenstates.
$G_{\psi K_{S, L}}, S_{\psi K_{S, L}}$ and $C_{\psi K_{S, L}}$
are parameters written in terms of $\lambda_{\psi K_{S, L}}$.
$\theta_{\psi K_{S, L}}, G_{\psi K_{S, L}}, S_{\psi K_{S, L}}$
and $C_{\psi K_{S, L}}$ explicitly appear in coefficients of the
master formula (\ref{M1}-\ref{M5})
for the processes.
In eq. (\ref{K1}), $\lambda_{\psi K_{S, L}}$ is written in terms of
the decay amplitude that the final state is the mass eigenstate $\psi K_{S, L}$.
$A_{\psi K_{S, L}}^{(\mathrm{ID})}$ and
$\bar{A}_{\psi K_{S, L}}^{(\mathrm{ID})}$ can be expanded with respect
to amplitudes with a flavor eigenstate as
$A_{\psi K^0}^{(\mathrm{ID})}, A_{\psi \bar{K^0}}^{(\mathrm{ID})},
\bar{A}_{\psi K^0}^{(\mathrm{ID})}$
and $\bar{A}_{\psi \bar{K^0}}^{(\mathrm{ID})}$.
The expressions of $A_{\psi K_{S, L}}^{(\mathrm{ID})}$
and $\bar{A}_{\psi K_{S, L}}^{(\mathrm{ID})}$
are exhibited in eqs. (\ref{D1}-\ref{D8}).
Note that the wrong strangeness decay amplitudes given as,
\bea
A_{\psi \bar{K^0}},\quad A_{\psi \bar{K^0}}^{\mathrm{ID}},\quad
\bar{A}_{\psi K^0}, \quad \bar{A}_{\psi K^0}^{\mathrm{ID}},
\eea
are numerically smaller than the right strangeness decay amplitudes
given as,
\bea
A_{\psi K^0},\quad A_{\psi K^0}^{\mathrm{ID}},\quad
\bar{A}_{\psi \bar{K^0}}, \quad
\bar{A}_{\psi \bar{K^0}}^{\mathrm{ID}},
\eea
for the standard model.
Thus, we ignore terms with higher power of
$A_{\psi \bar{K^0}}^{(\mathrm{ID})}$ and
$\bar{A}_{\psi K^0}^{(\mathrm{ID})}$.
Using eqs. (\ref{D1}-\ref{D8}), we can calculate
$\theta_{\psi K_{S, L}}$ as,
\bea
&\theta_{\psi K_S}\simeq \theta_K-z_K, \quad
\theta_{\psi K_L}\simeq \theta_K+z_K,\label{thzK}&\\
&\theta_K=
\displaystyle\frac{A_{\psi K^0}A_{\psi K^0}^{\mathrm{ID}}
-\bar{A}_{\psi \bar{K^0}}\bar{A}_{\psi \bar{K^0}}^{\mathrm{ID}}}
{A_{\psi K^0}A_{\psi K^0}^{\mathrm{ID}}
+\bar{A}_{\psi \bar{K^0}}\bar{A}_{\psi \bar{K^0}}^{\mathrm{ID}}},&
\eea
where
$\theta_K$ expresses CP and CPT violation in right strangeness
decays of B meson and it corresponds to $\hat{\theta}_{\psi K}$
of \cite{Applebaum:2013wxa}. Note that in
\cite{Applebaum:2013wxa} indirect CPT violation $z_K$
is not taken into account. 
When deriving eq. (\ref{thzK}), we calculated at linear order approximation
with respect to $z_K, \theta_K$ and wrong
strangeness decay amplitudes.
Then, we can write $\lambda_{\psi K_{S, L}}$ as,
\bea
&\lambda_{\psi K_{S}}\simeq\lambda(1-\Delta\lambda_{\mathrm{wst}}),\quad
\lambda_{\psi K_{L}}\simeq-\lambda(1+\Delta\lambda_{\mathrm{wst}}),
\label{K20}&\\
&\lambda\equiv
\displaystyle\frac{q}{p}
\displaystyle\frac{p_K}{q_K}
\displaystyle\frac{\bar{A}_{\psi \bar{K^0}}}{A_{\psi K^0}}
\sqrt{\displaystyle\frac{1+\theta_K}{1-\theta_K}}
=\displaystyle\frac{q}{p}
\displaystyle\frac{p_K}{q_K}
\displaystyle\frac{A_{\psi K^0}^{\mathrm{ID}}}
{\bar{A}_{\psi \bar{K^0}}^{\mathrm{ID}}}
\sqrt{\displaystyle\frac{1-\theta_K}{1+\theta_K}}
,\label{K7}&
\eea
where $\Delta\lambda_{\mathrm{wst}}$ consists of
wrong strangeness decays as,
\bea
&\Delta\lambda_{\mathrm{wst}}=
\lambda_{\psi \bar{K^0}}^{\mathrm{wst}}-\bar{\lambda}^{\mathrm{wst}}
_{\psi K^0},\label{A12pri}&
\\
&\lambda_{\psi \bar{K^0}}^{\mathrm{wst}}\equiv
\displaystyle\frac{p_K}{q_K}
\displaystyle\frac{A_{\psi\bar{K^0}}}{A_{\psi K^0}}
\sqrt{\displaystyle\frac{1+\theta_{\psi K^0}}{1-\theta_{\psi K^0}}}
=
\displaystyle\frac{p_K}{q_K}
\displaystyle\frac{\bar{A}_{\psi K^0}^{\mathrm{ID}}}
{\bar{A}_{\psi \bar{K^0}}^{\mathrm{ID}}}
\sqrt{\displaystyle\frac{1-\theta_{\psi K^0}}{1+\theta_{\psi K^0}}}
,\label{wst1}&\\
&\bar{\lambda}^{\mathrm{wst}}_{\psi K^0}\equiv
\displaystyle\frac{q_K}{p_K}
\displaystyle\frac{\bar{A}_{\psi K^0}}{\bar{A}_{\psi \bar{K^0}}}
\sqrt{\displaystyle\frac{1+\bar{\theta}_{\psi \bar{K^0}}}{1-\bar{\theta}_{\psi 
\bar{K^0}}}}
=
\displaystyle\frac{q_K}{p_K}
\displaystyle\frac{A^\mathrm{ID}_{\psi \bar{K^0}}}{A^\mathrm{ID}_{\psi K^0}}
\sqrt{\displaystyle\frac{1-\bar{\theta}_{\psi \bar{K^0}}}{1+\bar{\theta}_{\psi 
\bar{K^0}}}},\label{wst2}
\\
&\theta_{\psi K^0}\equiv
\displaystyle\frac{A_{\psi K^0}\bar{A}_{\psi K^0}^{\mathrm{ID}}-A_
{\psi \bar{K^0}}\bar{A}_{\psi \bar{K^0}}^{\mathrm{ID}}}
{A_{\psi K^0}\bar{A}_{\psi K^0}^{\mathrm{ID}}+A_
{\psi \bar{K^0}}\bar{A}_{\psi \bar{K^0}}^{\mathrm{ID}}},\quad
\bar{\theta}_{\psi \bar{K^0}}\equiv
\displaystyle\frac{\bar{A}_{\psi \bar{K^0}}A_{\psi 
\bar{K^0}}^{\mathrm{ID}}-\bar{A}_
{\psi K^0}A_{\psi K^0}^{\mathrm{ID}}}
{\bar{A}_{\psi \bar{K^0}}A_{\psi \bar{K^0}}^{\mathrm{ID}}+\bar{A}_
{\psi K^0}A_{\psi K^0}^{\mathrm{ID}}}\label{wstCPT},&
\eea
where eq. (\ref{wstCPT}) describes CPT violation in wrong strangeness
decays.
Similar to eq. (\ref{A12pri}), we can define a parameter including
wrong sign decay amplitudes as,
\bea
\hat{\lambda}_{\mathrm{wst}}=
\lambda_{\psi \bar{K^0}}^{\mathrm{wst}}+\bar{\lambda}^{\mathrm{wst}}
_{\psi K^0}
\label{hatwst}.
\eea
Since wrong sign semi-leptonic decay amplitudes and
CPT violation, $\theta_{\psi K^0}$ and $\bar{\theta}_{\psi \bar{K}^0}$,
are small, we can expand eqs. (\ref{wst1}, \ref{wst2}) as,
\bea
\lambda_{\psi \bar{K^0}}^{\mathrm{wst}}
\simeq
\displaystyle\frac{p_K}{q_K}\displaystyle\frac{A_{\psi \bar{K^0}}}{A_{\psi K^0}}
\simeq
\frac{p_K}{q_K}\frac{\bar{A}_{\psi K^0}^{\mathrm{ID}}}
{\bar{A}_{\psi \bar{K^0}}^{\mathrm{ID}}}
,\quad
\bar{\lambda}^{\mathrm{wst}}_{\psi K^0}\simeq
\displaystyle\frac{q_K}{p_K}
\displaystyle\frac{\bar{A}_{\psi K^0}}{\bar{A}_{\psi \bar{K^0}}}
\simeq
\displaystyle\frac{q_K}{p_K}\frac{A_{\psi \bar{K^0}}^{\mathrm{ID}}}{A_{\psi 
K^0}^{\mathrm{ID}}}
.\label{expandlambda}
\eea
Eq. (\ref{K20}) indicates that $\lambda_{\psi K_{S, L}}$ is composed of
the leading part $\lambda$
and the sub-leading part suppressed by wrong strangeness decay amplitude.
Taking the CPT conserving limit in eq. (\ref{K20}), one can obtain the relation in ref. \cite{Grossman:2002bu}.
Note that $\lambda$ has the definitive transformation property of
T, CP and CPT, such as $\lambda\xrightarrow{\mathrm{T}} (\lambda)^{-1}$,
$\lambda\xrightarrow{\mathrm{CP}} (\lambda)^{-1}$,
$\lambda\xrightarrow{\mathrm{CPT}} \lambda$.
One introduces $G, S$ and $C$ analogous to eq.(\ref{K1pppp})
by replacing $\lambda_{f}$
with $\lambda$.
They are transformed under T as,
\bea
\label{Gtra}G=
\displaystyle\frac{2\mathrm{Re}\lambda}{1+|\lambda|^2}
&\xrightarrow{\mathrm{T}}&
\displaystyle\frac{2\mathrm{Re}(1/\lambda)}{1+|1/\lambda|^2}
=\displaystyle\frac{2\mathrm{Re}\lambda^*}{|\lambda|^2+1}=+G,\\
\label{Stra}S=\displaystyle\frac{2\mathrm{Im}\lambda}{1+|\lambda|^2}
&\xrightarrow{\mathrm{T}}&
\displaystyle\frac{2\mathrm{Im}(1/\lambda)}{1+|1/\lambda|^2}
=\displaystyle\frac{2\mathrm{Im}\lambda^*}{|\lambda|^2+1}=-S,\\
\label{Ctra}C=\displaystyle\frac{1-|\lambda|^2}{1+|\lambda|^2}
&\xrightarrow{\mathrm{T}}&\displaystyle\frac{1-|1/\lambda|^2}{1+|1/\lambda|^2}
=\displaystyle\frac{|\lambda|^2-1}{|\lambda|^2+1}=-C.
\eea
The CP transformation property of $G, S, C$ is the same as (\ref{Gtra}-\ref{Ctra}).
Thus, the CPT transformation property is also determined as
$G\xrightarrow{\mathrm{CPT}}+G, S\xrightarrow{\mathrm{CPT}}+S,
C\xrightarrow{\mathrm{CPT}}+C$.
$|\lambda|$ is close to 1 since
deviation of $|q/p|, |p_K/q_K|$ and $|\bar{A}_{\psi \bar{K^0}}/A_{\psi K^0}|$ from 1 is small.
Hence, we can find that $C$ is a small parameter.
One can also derive the transformation property of eqs. (\ref{wst1}-\ref{wst2})
such like $\lambda_{\psi \bar{K^0}}^\mathrm{wst}\xrightarrow{\mathrm{T}}
\bar{\lambda}_{\psi K^0}^\mathrm{wst},
\lambda_{\psi \bar{K^0}}^\mathrm{wst}\xrightarrow{\mathrm{CP}}
\bar{\lambda}_{\psi K^0}^\mathrm{wst}$
and $\lambda_{\psi \bar{K^0}}^\mathrm{wst}\xrightarrow{\mathrm{CPT}}
\lambda_{\psi \bar{K^0}}^\mathrm{wst}$.
Therefore, the parameters in eqs. (\ref{A12pri}, \ref{hatwst}) are transformed
as,
\bea
\Delta \lambda_{\mathrm{wst}}\xrightarrow{\mathrm{T}}
-\Delta \lambda_{\mathrm{wst}},\quad
\hat{\lambda}_{\mathrm{wst}}\xrightarrow{\mathrm{T}}
\hat{\lambda}_{\mathrm{wst}}
\label{wstT}
.\eea
The CP transformation property of the parameter (\ref{A12pri}, \ref{hatwst})
is the same as eq. (\ref{wstT})
\\
Note that parameters
$G_{\psi K_{S, L}}, S_{\psi K_{S, L}}$ and
$C_{\psi K_{S, L}}$ are related with the parameters
$G, S$ and $C$ as,
\bea
&G_{\psi K_S}\simeq G+S\Delta\lambda_{\mathrm{wst}}^I
,\quad G_{\psi K_L}\simeq-( G-S\Delta\lambda_{\mathrm{wst}}^I),&\label{K8}\\
&S_{\psi K_S}\simeq S-G\Delta\lambda_{\mathrm{wst}}^I,
\quad S_{\psi K_L}\simeq-(S+G\Delta\lambda_{\mathrm{wst}}^I),&\label{K9}\\
&C_{\psi K_S}\simeq C+\Delta\lambda_{\mathrm{wst}}^R,
\quad C_{\psi K_L}\simeq C-\Delta\lambda_{\mathrm{wst}}^R.&
\label{K4}
\eea
where we use notation for an arbitrary complex number $A$,
$A^R\equiv\mathrm{Re}A, A^I\equiv\mathrm{Im}A$, throughout this paper.
When deriving eqs. (\ref{K8}-\ref{K4}), we ignored higher order terms of $C$ and $\Delta\lambda_{\mathrm{wst}}$.
Eqs. (\ref{thzK}, \ref{K8}-\ref{K4}) lead relations given as,
\bea
&\theta_{\psi K_S}+\theta_{\psi K_L}=2\theta_K,\label{K5}\quad
\theta_{\psi K_S}-\theta_{\psi K_L}= -2z_K,&\\
&G_{\psi K_{S}}-G_{\psi K_{L}}=2G,\quad
S_{\psi K_{S}}-S_{\psi K_{L}}=2S,\quad
C_{\psi K_{S}}+C_{\psi K_{L}}=2C,\label{K21}&\\
&G_{\psi K_{S}}+G_{\psi K_{L}}=2S\Delta \lambda^I_{\mathrm{wst}},\quad
S_{\psi K_{S}}+S_{\psi K_{L}}=-2G\Delta\lambda_{\mathrm{wst}}^I,\quad
C_{\psi K_{S}}-C_{\psi K_{L}}=2\Delta\lambda_{\mathrm{wst}}^R
.&\nn\\
\label{K22}
\eea
Since we have included the effect of indirect CP violation of Kaon system,
we show how the correction due to $\epsilon_K$ arises.
While the expression of $G,C$ and $S$ in eqs.(\ref{Gtra}-\ref{Ctra}) is invariant
under the arbitrary large rephasing such as 
$\bra{K^0} \rightarrow e^{-i \alpha_K} \bra{K^0}$
and $\bra{\bar{K^0}} \rightarrow e^{i \alpha_K} \bra{\bar{K^0}}$
, the parametrization with 
$\epsilon_K \ll 1$ allows only the small rephasing $\alpha_K \ll 1$.
\bea
\frac{p_K}{q_K}=\frac{1+\epsilon_K}{1-\epsilon_K}\simeq 1+ 2\epsilon_K. 
\eea 
Keeping only the terms which are linear to $\epsilon_K$, $G,S$ and $C$ 
are expanded as follows,
\bea
G&=&G^\prime-2 S^\prime \epsilon^I_K, \nn \\
S&=&S^\prime+2 G^\prime \epsilon^I_K, \nn \\
C&=&C^\prime-2 \epsilon^R_K, \label{eq:C}
\eea
where $G^\prime$, $S^\prime$ and $C^\prime$ are obtained by taking
the limit
$\frac{p_K}{q_K} \rightarrow 1$ in 
$G, S$ and $C$.
Namely, they are defined by replacing
$\lambda$ with $\lambda^\prime$
in the expression for $G, S$ and $C$.
\bea
\lambda^\prime=\frac{q}{p}\frac{\bar{A}_{\psi \bar{K^0}}}{A_{\psi K^0}}
\sqrt{\frac{1+\theta_K}{1-\theta_K}}, 
\quad
C^\prime=\frac{1-|\lambda^\prime|^2}{1+|\lambda^\prime|^2}\label{cpra2}
\eea
As shown in table(\ref{tab:1}), $(G^\prime, S^\prime, C^\prime)$ equal to 
$(\hat{G}_{\psi K}, \hat{S}_{\psi K}, \hat{C}_{\psi K})$ defined in
\cite{Applebaum:2013wxa} where indirect CP violation $\epsilon_K$ is 
neglected.
When one changes the phase convention of states as, 
the phase of $\lambda^\prime$ changes as follows,
\bea
&&\lambda^\prime \rightarrow \lambda^\prime e^{2 i \alpha_K}. 
\eea
Assuming the phase $\alpha_K$ is small,  $G^\prime$, $S^\prime$,
and $\epsilon^I_K$ change as, 
\bea
&&G^\prime \rightarrow G^\prime -2 \alpha_K S^\prime, \nn \\ 
&&S^\prime \rightarrow S^\prime + 2 \alpha_K G^\prime, \nn \\
&& \epsilon^I_K \rightarrow \epsilon^I_K-\alpha_K,
\eea
while $C^\prime$ and $\epsilon^R_K$ are invariant, 
\bea
C^\prime \rightarrow C^\prime, \quad \epsilon^R_K \rightarrow \epsilon^R_K.
\eea
Hereafter, we expand $C$ in terms of $C^\prime$ and $\epsilon^R_K$ as shown in
eq.(\ref{eq:C}) and we do not
expand $S$ and $G$ since they are invariant under the 
rephasing. The numerical significance of $\epsilon^R_K$ will be discussed in the next section. 
\\
We turn to definition for parameters including semi-leptonic decay
amplitudes of B meson. In the following, from eq.(\ref{Lep1}) to 
eq.(\ref{Lep5}), we adopt the notations of
\cite{Applebaum:2013wxa}.
Right sign semi-leptonic decay amplitudes are denoted as,
\bea
A_{l^+}=\braket{l^+X|T|B^0},\qquad
A_{l^+}^{\mathrm{ID}}=\braket{B^0|T|(l^+ X)^T},\nn\\
\bar{A}_{l^-}=\braket{l^-X|T|\bar{B^0}},\qquad
\bar{A}_{l^-}^{\mathrm{ID}}=\braket{\bar{B^0}|T|(l^- X)^T}
\label{Lep1}
.
\eea
Wrong sign semi-leptonic decay amplitudes are similarly given as,
\bea
A_{l^-}=\braket{l^-X|T|B^0},\qquad
A_{l^-}^{\mathrm{ID}}=\braket{B^0|T|(l^-X)^T},\nn\\
\bar{A}_{l^+}=\braket{l^+X|T|\bar{B^0}},\qquad
\bar{A}_{l^+}^{\mathrm{ID}}=\braket{\bar{B^0}|T|(l^+ X)^T}
\label{Lep2}
.
\eea
For the case of the standard model,
wrong sign semi-leptonic decay amplitudes are smaller than
right sign decay amplitudes.
Thus, we ignore higher powers of wrong sign decay amplitudes
than linear order.
We define parameters including semi-leptonic decay amplitudes
as,
\bea
\lambda_{l^+}&\equiv&\frac{q}{p}\frac{\bar{A}_{l^+}}{A_{l^+}}
\sqrt{\frac{1+\theta_{l^+}}{1-\theta_{l^+}}}=
\frac{q}{p}\frac{A_{l^-}^{\mathrm{ID}}}{\bar{A}_{l^-}^{\mathrm{ID}}}
\sqrt{\frac{1-\theta_{l^+}}{1+\theta_{l^+}}},\quad
\theta_{l^+}=
\frac{A_{l^+}A_{l^-}^{\mathrm{ID}}-\bar{A}_{l^+}\bar{A}_{l^-}^{\mathrm{ID}}}
{A_{l^+}A_{l^-}^{\mathrm{ID}}+\bar{A}_{l^+}\bar{A}_{l^-}^{\mathrm{ID}}},
\label{Lep3}\\
\lambda_{l^-}&\equiv&\frac{q}{p}\frac{\bar{A}_{l^-}}{A_{l^-}}
\sqrt{\frac{1+\theta_{l^-}}{1-\theta_{l^-}}}=
\frac{q}{p}\frac{A_{l^+}^{\mathrm{ID}}}{\bar{A}_{l^+}^{\mathrm{ID}}}
\sqrt{\frac{1-\theta_{l^-}}{1+\theta_{l^-}}}
,\quad
\theta_{l^-}=\frac{A_{l^-}A_{l^+}^{\mathrm{ID}}-\bar{A}_{l^-}\bar{A}_{l^+}^{\mathrm{ID}}}
{A_{l^-}A_{l^+}^{\mathrm{ID}}-\bar{A}_{l^-}\bar{A}_{l^+}^{\mathrm{ID}}}
\label{Lep4},\eea
where $\theta_{l^\pm}$ expresses CPT violation in semi-leptonic decays
of B meson.
One can find the transformation law such like $\lambda_{l^+}\xrightarrow{T} (\lambda_{l^-})^{-1}$,
$\lambda_{l^+}\xrightarrow{CP} (\lambda_{l^-})^{-1}$,
$\lambda_{l^+}\xrightarrow{CPT} \lambda_{l^+}$
by its definition (\ref{Lep3}-\ref{Lep4}).
We assume that CPT violating parameter $\theta_{l^\pm}$
is small.
At linear order approximation of $\theta_{l^\pm}$ and wrong sign
semi-leptonic decay amplitudes,
we obtain,
\bea
\lambda_{l^+}\simeq\frac{q}{p}\frac{\bar{A}_{l^+}}{A_{l^+}}
\simeq
\frac{q}{p}\frac{A_{l^-}^{\mathrm{ID}}}{\bar{A}_{l^-}^{\mathrm{ID}}}
,\quad
\lambda_{l^-}^{-1}\simeq\frac{p}{q}\frac{A_{l^-}}{\bar{A}_{l^-}}
\simeq\frac{p}{q}\frac{\bar{A}_{l^+}^\mathrm{ID}}{A_{l^+}^{\mathrm{ID}}},
\label{Lep5}
\eea
where we can see that contribution of $\theta_{l^\pm}$ approximately
vanishes in eq. (\ref{Lep5}).
Following ref.\cite{Applebaum:2013wxa}, we also define $G_{l^\pm}, S_{l^\pm}$ and $C_{l^\pm}$ analogous to eq.(\ref{K1pppp}) by replacing $\lambda_f$ with $\lambda_{l^\pm}$.
The parameters
$G_{l^\pm}, S_{l^\pm}$ and $C_{l^\pm}$
explicitly appear in
coefficients of master formula (\ref{M1}-\ref{M5})
for the processes in which
the final states are given as
$l^\pm X$.
Eq.(\ref{Lep5}) gives approximate expressions for $G_{l^\pm}, S_{l^\pm}$
and $C_{l^\pm}$ as,
\bea
&G_{l^+}=\displaystyle\frac{2\mathrm{Re}\lambda_{l^+}}{1+|\lambda_{l^+}|^2}
\simeq 2\mathrm{Re}\lambda_{l^+},\:
G_{l^-}=\displaystyle\frac{2\mathrm{Re}\lambda_{l^-}}{1+|\lambda_{l^-}|^2}\simeq 2\mathrm{Re}(\lambda_{l^-}^{-1}),\:
C_{l^\pm}=\displaystyle\frac{1-|\lambda_{l^\pm}|^2}{1+|\lambda_{l^\pm}|^2}
\simeq \pm 1,&
\nn\\
&S_{l^+}=\displaystyle\frac{2\mathrm{Im}\lambda_{l^+}}{1+|\lambda_{l^+}|^2}\simeq 2\mathrm{Im}\lambda_{l^+},\:\:
S_{l^-}=\displaystyle\frac{2\mathrm{Im}\lambda_{l^-}}{1+|\lambda_{l^-}|^2}
\simeq -2\mathrm{Im}(\lambda_{l^-}^{-1})
\label{Lep8}
.&
\eea
Note that eq. (\ref{Lep8}) implies that
$G_{l^\pm}$ and $S_{l^\pm}$ are small numbers since
$\lambda_{l^+}$ and $\lambda_{l^-}^{-1}$ are suppressed.
We can find the relations,
\bea
&G_{l^+}+G_{l^-}=2\hat{\lambda}_l^R,\quad
S_{l^+}-S_{l^-}=2\hat{\lambda}_l^I\label{Lep9},&\\
&G_{l^+}-G_{l^-}=2\Delta\lambda_l^R,\quad
S_{l^+}+S_{l^-}=2\Delta\lambda_l^I\label{Lep10},&
\eea
where $\hat{\lambda}_l$ and $\Delta\lambda_l$ are defined as,
\bea
\hat{\lambda}_l\equiv \lambda_{l^+}+\lambda_{l^-}^{-1},\quad
\Delta\lambda_l\equiv\lambda_{l^+}-\lambda_{l^-}^{-1}.
\eea
They transform definitively under CP, T and CPT.
One obtains the transformation property of T as,
\bea
\hat{\lambda}_l\xrightarrow{\mathrm{T}}(\lambda_{l^-})^{-1}+\lambda_{l^+}
=+\hat{\lambda}_l,\quad
\Delta\lambda_{l}\xrightarrow{\mathrm{T}}(\lambda_{l^-})^{-1}-\lambda_{l^+}=-\Delta\lambda_{l}.
\label{Tl}
\eea
The CP transformation property of $\hat{\lambda}_l$ and
$\Delta\lambda_{l}$ is the same as (\ref{Tl}).
Hence, the CPT transformation property of $\hat{\lambda}_l$ and
$\Delta\lambda_{l}$ is also determined as $\hat{\lambda}_l\xrightarrow
{\mathrm{CPT}}\hat{\lambda}_l, \Delta\lambda_l\xrightarrow
{\mathrm{CPT}}\Delta\lambda_l$.
Eqs. (\ref{K8}-\ref{K4}, \ref{Lep9}-\ref{Lep10}) enable one to write down
the asymmetry in eq.(\ref{T2}) for the BaBar experiment in terms of parameters which are exactly T-odd or T-even.
Similarly, one defines,
\bea
&R_M\equiv
\displaystyle\frac{|p|^2-|q|^2}{|p|^2+|q|^2}
,\quad
\xi_l\equiv
\displaystyle\frac{\bar{A}_{l^-}A_{l^+}^{\mathrm{ID}}-A_{l^+}\bar{A}_{l^-}^{\mathrm{ID}}}
{\bar{A}_{l^-}A_{l^+}^{\mathrm{ID}}+A_{l^+}\bar{A}_{l^-}^{\mathrm{ID}}},\quad
C_\xi^l,\equiv
\displaystyle\frac{1-|\lambda_\xi^l|^2}{1+|\lambda_\xi^l|^2}&
\label{paraxi}\\
&\lambda^l_\xi\equiv
\displaystyle\frac{A_{l^+}}{\bar{A}_{l^-}}\sqrt{\frac{1+\xi_l}{1-\xi_l}}
=\displaystyle\frac{A_{l^+}^{\mathrm{ID}}}
{\bar{A}_{l^-}^{\mathrm{ID}}}\sqrt{\frac{1-\xi_{l}}
{1+\xi_{l}}}.\label{lambdaxi}
&
\eea
In eq. (\ref{paraxi}), $R_M$ 
expresses mixing-induced CP and T violation
for B meson system \cite{Applebaum:2013wxa} and is a small number.
As for newly introduced parameters,
$\xi_l$ implies CP and T violation in right sign semi-leptonic decays
and we also assume $\xi_l$ is a small number.
The expression of $\lambda_\xi^l$ (\ref{lambdaxi}) includes
right sign semi-leptonic decay amplitude ratio
and we assume $|A_{l^+}/\bar{A}_{l^-}|\simeq 1$.
Therefore, $C_{\xi}^l$ is a small number compared with $\mathcal{O}(1)$.
The parameters defined in eq. (\ref{paraxi}) also appear in 
the asymmetry in eq.(\ref{T2}).
\par
Hereafter, we describe some significant points of the parameters
defined in this section.
Note that the parameters given as,
\bea
S, C, G, \theta_K, R_M, z, z_K, \hat{\lambda}_l, \Delta \lambda_l, \xi_l,
C_{\xi}^l, \hat{\lambda}_{\mathrm{wst}}\:\mathrm{and}\:\Delta\lambda_{\mathrm{wst}},
\label{defi}
\eea
have definitive transformation properties exhibited
in table \ref{Ta1}.
In the
processes which we consider,
 $K_{S, L}$ is included as a final state,
and the effect of mixing induced T and CP
violation, $p_K/q_K$, appears in the expressions of $G, S, C, 
\hat{\lambda}_{\mathrm{wst}}$ and $\Delta\lambda_{\mathrm{wst}}$.
Mixing-induced CP and CPT violation in K meson system, $z_K$,
reveals in the asymmetry as well.
In the next section, 
the asymmetry is written in terms of parameters (\ref{defi})
and it can be explicitly separated as T-odd parts and T-even parts.\\
The parameters defined as,
\bea
p/q, p_K/q_K, \theta_{\psi K^0}, \bar{\theta}_{\psi \bar{K^0}},
\theta_{l^\pm},
\lambda, \lambda_{\psi\bar{K^0}}^{\mathrm{wst}},
\bar{\lambda}_{\psi K^0}^{\mathrm{wst}},
\lambda_{l^\pm},
\:\mathrm{and}\:\lambda_\xi^l,\label{devoted}
\eea
are dedicated to keep the definitive transformation property of parameters that reveal in table \ref{Ta1}.
The transformation property of the parameters (\ref{devoted}) is exhibited in
table \ref{Ta2}.\\
The parameters given as,
\bea
\theta_{\psi K^0}, \bar{\theta}_{\psi \bar{K^0}}, \theta_{l^\pm},
C, \theta_K, R_M, z, z_K, \hat{\lambda}_l, \Delta\lambda_l, \xi_l, C_\xi^l,
\hat{\lambda}_{\mathrm{wst}}\:\mathrm{and}\:\Delta\lambda_{\mathrm{wst}},
\label{sma}
\eea
are small numbers, and our calculation is based
on linear order approximation with respect to the parameters (\ref{sma})
throughout this paper.
\begin{table}[hbtp]
 \caption{Transformation property of the parameters definitively transformed under T, CP and CPT}
 \label{Ta1}
 \begin{center}
  \begin{tabular}{|c||c|c|c|c|c|c|c|c|c|c|c|c|c|}
   \hline
     &$S$&$C$&$G$&$\theta_K$&$R_M$& $z$&$z_K$&$\hat{\lambda}_l$&$\Delta\lambda_l$  &$\xi_l$ &  $C_\xi^l$&$\hat{\lambda}_{\mathrm{wst}}$ &$\Delta\lambda_{\mathrm{wst}}$\\
   \hline \hline
   T  &$-$&$-$&$+$&$+$&$-$&$+$&$+$&$+$&$-$ &$-$  & $+$&$+$&$-$\\\hline
   CP  &$-$&$-$&$+$&$-$&$-$&$-$&$-$&$+$&$-$ &$-$   & $-$&$+$&$-$\\\hline
   CPT  &$+$&$+$&$+$&$-$&$+$&$-$&$-$&$+$&$+$ &$+$  & $-$&$+$&$+$\\
   \hline
  \end{tabular}
 \end{center}
\end{table}
\begin{table}[hbtp]
 \caption{Transformation property of the parameters devoted to keep 
the definitive transformation property of the parameters in table\ref{Ta1}}
 \label{Ta2}
 \begin{center}
  \begin{tabular}{|c||c|c|c|c|c|c|c|c|c|}
   \hline
     &$p/q$&$p_K/q_K$&$\theta_{\psi K^0}$&$\theta_{l^+}$&$\lambda$&
     $\lambda_{\psi \bar{K^0}}^{\mathrm{wst}}$&
     $\lambda_{l^+}$&$\lambda_\xi^l$\\
   \hline \hline
   T  &$q/p$&$q_K/p_K$&$-\bar{\theta}_{\psi \bar{K^0}}$&$\theta_{l^-}$&$(\lambda)^{-1}$&
   $\bar{\lambda}^{\mathrm{wst}}_{\psi K^0}$&
   $(\lambda_{l^-})^{-1}$&$\lambda_\xi^l$\\\hline
   CP  &$q/p$&$q_K/p_K$&$\bar{\theta}_{\psi \bar{K^0}}$&$-\theta_{l^-}$&$(\lambda)^{-1}$&
   $\bar{\lambda}^{\mathrm{wst}}_{\psi K^0}$&
   $(\lambda_{l^-})^{-1}$&$(\lambda_\xi^l)^{-1}$\\\hline
   CPT  &$p/q$&$p_K/q_K$&$-\theta_{\psi K^0}$&$-\theta_{l^+}$&$\lambda$&
   $\lambda_{\psi \bar{K^0}}^{\mathrm{wst}}$&
   $\lambda_{l^+}$&$(\lambda_\xi^l)^{-1}$\\
   \hline
  \end{tabular}
 \end{center}
\end{table}
\section{Time dependent asymmetry including the overall constants}
\label{Sec:2}
In this section, we apply the event number asymmetry defined in eq.(\ref{T6}) to
processes for B-meson decays.
One should be aware of that
the asymmetry considered
in this paper includes the effect
of different normalization for
two rates; non-zero value of
$\Delta N_R$ defined in eq.(\ref{T7}).
As the BaBar asymmetry investigated in
\cite{Lees:2012uka},
we also assign $f_1, f_2, f_3, f_4$ with $\psi K_L, l^-X,$ $l^+X, \psi K_S$, respectively.  We call this process as ${\rm I}$.
We also consider the other three processes which can be obtained by
interchanging $l^- X$ with $l^+ X$ and $\psi K_S$ with $\psi K_L$ in the 
process ${\rm I}$.
Therefore we identify the four processes as,
\bea
({\rm I}) && (f_1,f_2,f_3,f_4)=(\psi K_L, l^-X, l^+X, \psi K_S), \nn \\
({\rm II}) &&(f_1,f_2,f_3,f_4)=(\psi K_S, l^-X, l^+X, \psi K_L), \nn \\
({\rm III})&&(f_1,f_2,f_3,f_4)=(\psi K_L, l^+X, l^-X, \psi K_S), \nn \\
({\rm IV}) && (f_1,f_2,f_3,f_4)=(\psi K_S, l^+X, l^-X, \psi K_L).
\label{Bprocesses}
\eea
For all the processes which we consider,
we can find $\Delta N_R, \Delta \sigma, y\Gamma t,
\Delta \C, \hat{\S}, \hat{\C}$ are small numbers compared with $\mathcal{O}(1)$. Expanding eq. (\ref{T6}) with respect to the 
small parameters, one obtains the
asymmetry at linear order
approximation,
\bea
A &\simeq&
R_T+C_T\cos(x\Gamma t)+S_T\sin(x\Gamma t)\nn\\
&&+B_T\sin^2(x\Gamma t)
+D_T\sin(x\Gamma t)\cos(x\Gamma t)+E_T (y\Gamma t)
\sin(x\Gamma t),
\label{T8}
\eea
where,
\bea
R_T&=&-\frac{\Delta N_R}{2}+\frac{\Delta \sigma}{2}y\Gamma t
\simeq -\frac{\Delta N_R}{2} \label{RT},\\
C_T&=&\frac{\Delta \C}{2}, \quad 
S_T=\frac{\Delta \S}{2},
\label{ST}\\
B_T&=&-\frac{\Delta \S}{4}\hat{S}, \quad
D_T=-\frac{\Delta \S}{4}\hat{C},  \label{BT}\\
E_T&=&-\frac{\Delta\mathcal{S}}{4}\hat{\sigma}.\label{ET}
\eea
We ignore $\Delta\sigma y$ in eqs. (\ref{RT}-\ref{ET}).
$\hat{\sigma}$ and $\Delta\mathcal{S}$ are $\mathcal{O}(1)$ parameters
 and $\hat{\sigma} y$ gives rise to small contribution.
The model independent parametrization in eq. (\ref{T8}) without the last term 
can be found in \cite{Applebaum:2013wxa}.
In each process, we compute the asymmetry and the coefficients $(R_T, C_T, S_T, B_T, D_T, E_T)$. We label suffix ${\rm I} \sim {\rm IV}$ on the quantities corresponding to each process to distinguish them.
Below and in table \ref{Ta3}, 
we show the asymmetry and the coefficients for the process ${\rm I}$. 
For the  other processes, we show them in tables \ref{Tab4}-\ref{Tab6}.
 We first investigate $\Delta N_R$ in eq. (\ref{T7}) for the process ${\rm I}$.
With eq. (\ref{NR}),
one obtains,
\bea
\Delta N^{I}_R=2[-Sz^I+R_M+\hat{\lambda}_{\mathrm{wst}}^R-G
\hat{\lambda}_l^{R}-C^l_\xi-\xi_l^R].\label{DNR}
\eea
With eq. (\ref{DNR}) and eqs. (\ref{delS})-(\ref{A12}),
one can also derive the coefficients in the asymmetry,
\bea
R^{I}_T&=&-\frac{\Delta N_R^I}{2}
=
Sz^I-R_M-\hat{\lambda}_{\mathrm{wst}}^R+G
\hat{\lambda}_l^{R}+C^l_\xi+\xi_l^R,
\label{T9}\\
C^{I}_T&=&\frac{\Delta \C^I}{2}=C-Sz^I +\theta_K^R+ S\Delta\lambda_l^I
=C^\prime-2\epsilon_K^R-Sz^I+\theta_K^R+S\Delta\lambda_l^I,
\label{T10}\\
S^{I}_T&=&\frac{\Delta \S^I}{2}
=
-[S(1-Gz^R)-G\theta_K^I+GS\Delta\lambda_l^R],
\\
B^{I}_T&=&-\frac{\Delta \S^I}{4}\hat{S}^I\simeq \frac{S}{2}\hat{S}^I\nn\\
&=&S[G(z_K^I-\Delta\lambda_{\mathrm{wst}}^I)-z^I
+SR_M
+S\hat{\lambda}_{\mathrm{wst}}^R-SC_\xi^l-S\xi_l^R],\label{T14prime}\\
D^{I}_T&=&-\frac{\Delta \S^I}{4}\hat{C}^I\simeq \frac{S}{2}\hat{C}^I
=S[z_K^R-\Delta\lambda_{\mathrm{wst}}^R-Gz^R-S\hat{\lambda_l^I}],\label{T13}\\
E^{I}_T&=&-\frac{\Delta\mathcal{S}^I}{4}\hat{\sigma}^I\simeq GS.\label{T14}
\eea 
Note that $C^\prime$ and $\epsilon_K^R$
are phase convention independent parameters
due to definition of $C^\prime$.
Therefore, we state that all of eqs. (\ref{T9}-\ref{T14})
are expressed as phase convention independent parameters.
In eq. (\ref{T10}), effect of indirect CP violation in K meson
system explicitly appears
and gives rise to $\mathcal{O}(10^{-3})$ contribution to
$C_T^I$.
Assuming $|q/p|-1, |\bar{A}_{\psi \bar{K^0}}/A_{\psi K^0}|-1$,
$|1+\theta_K|-1$ are small numbers,
we can expand $C^\prime$ in eq.(\ref{cpra2}) as,
\bea
C^\prime\simeq
2-\left|\frac{q}{p}\right|-\left|\frac{\bar{A}_{\psi \bar{K^0}}}{A_{\psi K^0}}\right|-\theta_K^R,\quad
\left|\frac{q}{p}\right|\simeq
1-\frac{1}{2}\mathrm{Im}\left(\frac{\Gamma_{12}^d}{M^d_{12}}
\right)
\label{Cprasim}
\eea
A theoretical prediction for the imaginary part of
$\Gamma_{12}^d/M_{12}^d$ is calculated \cite{Lenz:2011ti},
and it shows $\mathrm{Im}(\Gamma_{12}^d/M_{12}^d)
\sim \mathcal{O}(10^{-4})$.
Direct CP violation in $B^0_d\rightarrow \psi K^0$
is $1-|\bar{A}_{\psi \bar{K^0}}/A_{\psi K^0}|\simeq
\mathcal{O}(10^{-3})$\cite{Bigi:2000yz}-\cite{Li:2006vq}.
Hence, $\epsilon_K^R, |\bar{A}_{\psi \bar{K^0}}/A_{\psi K^0}|\sim \mathcal{O}(10^{-3})$
are dominant in $C_T^I$,
if CPT violations
and the wrong sign decay in $B\rightarrow l X$ in eq. (\ref{T10}) are also negligible.
\par
If $R_T, C_T, S_T, B_T$ and $D_T$ were genuine T-odd quantities, they would vanish in the limit
of T-symmetry.
In other words, if there are non-vanishing contributions in the
limit of T-symmetry,
$R_T, C_T, S_T, B_T$ and $D_T$
are not T-odd quantities.
From eqs.(\ref{T9}-\ref{T14}),
we find the T-even contributions.
Some of them do not vanish in the limit of T-symmetry and they
include $C_\xi^l, \hat{\lambda}_{\mathrm{wst}}^R
\theta_K^R$, etc.
The others are quadratic with respect to
T-odd quantities and
they vanish in the limit of T-symmetry. They include 
$S\Delta\lambda_l^I, S\Delta\lambda_{\mathrm{wst}}^R,
S^2\hat{\lambda}_l^I,$ etc.
\par
Now we study condition that the asymmetry becomes a T-odd quantity.
The following equations are needed for T-even terms in each coefficient to vanish,
\bea
\hat{\lambda}_{\mathrm{wst}}^R =0, \quad G \hat{\lambda}_l^{R}=0,\quad 
 C^l_\xi=0 &\rightarrow&   R_T^I:\mathrm{T}-\mathrm{odd}, \label{Todd1}\\
 \theta_K^R=0, \quad S\Delta\lambda_l^I=0 &\rightarrow&  C_T^I:\mathrm{T}-\mathrm{odd},  \\
 G\theta_K^I=0, \quad GS\Delta\lambda_l^R=0  &\rightarrow&  S_T^I:\mathrm{T}-\mathrm{odd}, \label{Todd2} \\
 SG\Delta\lambda_{\mathrm{wst}}^I=0, \quad S^2\hat{\lambda}_{\mathrm{wst}}^R=0, \quad S^2C_\xi^l=0  
&\rightarrow&  B_T^I: \mathrm{T}-\mathrm{odd}, \label{Todd3} \\
 S\Delta\lambda_{\mathrm{wst}}^R=0,\quad  S^2\hat{\lambda_l^I}=0 &\rightarrow& 
  D_T^I:\mathrm{T}-\mathrm{odd}. 
\label{Todd4}
 \eea
When the real part and imaginary part of $\lambda$ do not vanish,
both $G$ and $S$ are non-zero and 
the conditions that all the eqs. (\ref{Todd1})-(\ref{Todd4}) 
are satisfied become, 
\bea
\theta_K=\Delta\lambda_{\mathrm{wst}}=\Delta\lambda_l=\hat{\lambda}_l=\hat{\lambda}_{\mathrm{wst}}^R=C^l_\xi=0.
\label{eq:parameterzero}
\eea
The conditions except $C^l_{\xi}=0$ agree with ones obtained
in \cite{Applebaum:2013wxa}. The additional condition is 
required since we take account of the overall constants in the asymmetry.
\par
In the first column of table \ref{Ta3}, 
we show how each coefficient of the asymmetry in eq.(\ref{T8}) depends on T-odd combination
of the parameters and in the other columns we show the dependence on T-even combination of the parameters.
As for T-even contribution, we identify the sources of T-even contribution to the coefficients.
In the second column, the contribution of
$\theta_K$ which is  CP and CPT violation in right strangeness decays is shown. In the third column, the contribution 
of $C^l_{\xi}$ which is CP and CPT violation in the right sign semi-leptonic decays 
is shown.  In the fourth and the fifth column, T-even contribution from
the wrong strangeness decays and the wrong sign semi-leptonic decays are shown, respectively.
\begin{table}[hbtp]
 \caption{
The coefficients of the asymmetry for the process I with the final state
$(f_1, f_2, f_3, f_4)=(\psi K_L, l^-X, l^+X, \psi K_S)$
and the sources which give
rise to the non-vanishing contribution to the asymmetry. The sources of the first column corresponds to
T-odd terms and the other correspond to T-even terms.
}
 \label{Ta3}
 \begin{center}
  \begin{tabular}{|c||c|c|c|c|c|}
   \hline
     $$&T-odd terms&$\theta_K \neq0$&$C^l_\xi \neq0$&$A_{\psi \bar{K^0}}\neq0,\bar{A}_{\psi K^0}\neq0$&$\bar{A}_{l^+}\neq0,A_{l^-}\neq0$\\\hline \hline
   $R^I_T$&$Sz^I-R_M+\xi_l^R $ &$0$&$C_{\xi}^l$&$-\hat{\lambda}_{\mathrm{wst}}^R$&$G\hat{\lambda}_l^R$\\ \hline
   $C^I_T$&$C-Sz^I$  &$\theta_K^R$&$0$&$0$&$S\Delta\lambda_l^I$\\\hline
   $S^I_T$&$-S[1-Gz^R]$  &$G\theta_K^I$&$0$&$0$&$-GS\Delta\lambda_l^R$\\\hline
   $B^I_T$&$S[Gz_K^I-z^I+SR_M-S\xi_l^R]$  &$0$&$-S^2C_{\xi}^l$&$S^2\hat{\lambda}_{\mathrm{wst}}^R-SG\Delta\lambda_{\mathrm{wst}}^I$&$0$\\\hline
   $D^I_T$&$S[z_K^R-Gz^R]$  &$0$&$0$&$-S\Delta\lambda_{\mathrm{wst}}^R$&$-S^2\hat{\lambda}_l^I$\\\hline
   $E^I_T$&$GS$  &$0$&$0$&$0$&$0$\\
   \hline
  \end{tabular}
 \end{center}
\label{Tab3}
\end{table}
\begin{table}[hbtp]
 \caption{The coefficients of the asymmetry for the process II with the final state $(f_1, f_2, f_3, f_4)=(\psi K_S, l^-X, l^+X, \psi K_L) $
and the sources which give
rise to the non-vanishing contribution to the asymmetry. The sources of the first column corresponds to
T-odd terms and the other correspond to T-even terms.
}
 \label{Tax}
 \begin{center}
  \begin{tabular}{|c||c|c|c|c|c|}
   \hline
     $$&T-odd terms&$\theta_K \neq0$&$C^l_\xi \neq0$&$A_{\psi \bar{K^0}}\neq0,\bar{A}_{\psi K^0}\neq0$&$\bar{A}_{l^+}\neq0,A_{l^-}\neq0$\\\hline \hline
   $R^{II}_T$&$-Sz^I-R_M+\xi_l^R $ &$0$&$C_{\xi}^l$&$\hat{\lambda}_{\mathrm{wst}}^R$&$-G\hat{\lambda}_l^R$\\\hline
   $C^{II}_T$&$C+Sz^I$  &$\theta_K^R$&$0$&$0$&$-S\Delta\lambda_l^I$\\\hline
   $S^{II}_T$&$S[1+Gz^R]$  &$-G\theta_K^I$&$0$&$0$&$-GS\Delta\lambda_l^R$\\\hline
   $B^{II}_T$&$-S[Gz_K^I-z^I-SR_M+S\xi_l^R]$  &$0$&$-S^2C_{\xi}^l$&$-S^2\hat{\lambda}_{\mathrm{wst}}^R+SG\Delta\lambda_{\mathrm{wst}}^I$&$0$\\\hline
   $D^{II}_T$&$S[z_K^R-Gz^R]$  &$0$&$0$&$-S\Delta\lambda_{\mathrm{wst}}^R$&$-S^2\hat{\lambda}_l^I$\\\hline
   $E^{II}_T$&$GS$  &$0$&$0$&$0$&$0$\\
   \hline
  \end{tabular}
 \end{center}
\label{Tab4}
\end{table}
\begin{table}[hbtp]
\caption{The coefficients of the asymmetry for the process III with 
the final states 
$(f_1, f_2, f_3, f_4)=(\psi K_L, l^+X, l^-X, \psi K_S)$
and the sources which give
rise to the non-vanishing contribution to the asymmetry. The sources of the first column corresponds to
T-odd terms and the other correspond to T-even terms.
} 
\begin{center}
  \begin{tabular}{|c||c|c|c|c|c|}
   \hline
   $$& T-odd terms&$\theta_K \neq0$&$C^l_\xi \neq0$&$A_{\psi \bar{K^0}}\neq0,\bar{A}_{\psi K^0}\neq0$&$\bar{A}_{l^+}\neq0,A_{l^-}\neq0$\\\hline \hline
   $R^{III}_T$&$Sz^I+R_M-\xi_l^R $ &$0$&$-C_{\xi}^l$&$-\hat{\lambda}_{\mathrm{wst}}^R$&$G\hat{\lambda}_l^R$\\\hline
   $C^{III}_T$&$-C-Sz^I$  &$-\theta_K^R$&$0$&$0$&$S\Delta\lambda_l^I$\\\hline
   $S^{III}_T$&$S[1+Gz^R]$  &$-G\theta_K^I$&$0$&$0$&$-GS\Delta\lambda_l^R$\\\hline
   $B^{III}_T$&$S[Gz_K^I-z^I-SR_M+S\xi_l^R]$  &$0$&$S^2C_{\xi}^l$&$S^2\hat{\lambda}_{\mathrm{wst}}^R-SG\Delta\lambda_{\mathrm{wst}}^I$&$0$\\\hline
   $D^{III}_T$&$S[z_K^R-Gz^R]$  &$0$&$0$&$-S\Delta\lambda_{\mathrm{wst}}^R$&$-S^2\hat{\lambda}_l^I$\\\hline
   $E^{III}_T$&$-GS$  &$0$&$0$&$0$&$0$\\
   \hline
  \end{tabular}
 \end{center}
\label{Tab5}
\end{table}
 \begin{table}[hbtp]
\caption{The coefficients of the asymmetry for the process IV with 
the final states 
$(f_1, f_2, f_3, f_4)=(\psi K_S, l^+X, l^-X, \psi K_L)$
and the sources which give
rise to the non-vanishing contribution to the asymmetry. The sources of the first column corresponds to
T-odd terms and the other correspond to T-even terms.
} 
\begin{center}
  \begin{tabular}{|c||c|c|c|c|c|}
   \hline
   $$& T-odd terms&$\theta_K \neq0$&$C^l_\xi \neq0$&$A_{\psi \bar{K^0}}\neq0,\bar{A}_{\psi K^0}\neq0$&$\bar{A}_{l^+}\neq0,A_{l^-}\neq0$\\\hline \hline
   $R^{IV}_T$&$-Sz^I+R_M-\xi_l^R $ &$0$&$-C_{\xi}^l$&$\hat{\lambda}_{\mathrm{wst}}^R$&$-G\hat{\lambda}_l^R$\\\hline
   $C^{IV}_T$&$-C+Sz^I$  &$-\theta_K^R$&$0$&$0$&$-S\Delta\lambda_l^I$\\\hline
   $S^{IV}_T$&$-S[1-Gz^R]$  &$G\theta_K^I$&$0$&$0$&$-GS\Delta\lambda_l^R$
\\\hline
   $B^{IV}_T$&$S[-Gz_K^I+z^I-SR_M+S\xi_l^R]$  &$0$&$S^2C_{\xi}^l$&$-S^2\hat{\lambda}_{\mathrm{wst}}^R+SG\Delta\lambda_{\mathrm{wst}}^I$&$0$\\\hline
   $D^{IV}_T$&$S[z_K^R-Gz^R]$  &$0$&$0$&$-S\Delta\lambda_{\mathrm{wst}}^R$&$-S^2\hat{\lambda}_l^I$\\\hline
   $E^{IV}_T$&$-GS$  &$0$&$0$&$0$&$0$\\
   \hline
  \end{tabular}
 \end{center}
\label{Tab6}
\end{table}
In tables \ref{Tab4}-\ref{Tab6}, 
we show the coefficients for the processes (II)-(IV).
In appendix \ref{F}, we show a rule useful for deriving them.
\begin{table}
\caption{Combinations of the independent coefficients in the asymmetry.
The sources which contribute to each combination are classified
in three categories.}
\begin{center}
\begin{tabular}{|c|c|c|c|} \hline
 &CPT even parameters  & CPT violating parameters & wrong sign decays \\ \hline
$\frac{R_T^I+R_T^{II}}{2}$&$-R_M+\xi_l^R$& $+C_\xi^l$ & 0\\ \hline
$\frac{R_T^I-R_T^{II}}{2}$&$0$ & $S z^I$ & $ -\hat{\lambda}_{\mathrm{wst}}^R+G\hat{\lambda}_l^R$  \\ \hline
$\frac{C_T^I+C_T^{II}}{2}$&$C$& $\theta_K^R$& $0$ \\ \hline
$\frac{C_T^I-C_T^{II}}{2}$&$0$&$-S z_I$ &$S \Delta \lambda_l^I$ \\ \hline
$\frac{S_T^I+S_T^{II}}{2}$&$0$& $S G z^R$& $-S G \Delta \lambda_l^R$  \\ \hline
$\frac{S_T^I-S_T^{II}}{2}$&$-S$&$G \theta_K^I$ & $0$ \\ \hline
$\frac{B_T^I+B_T^{II}}{2}$&$S^2(R_M-\xi_l^R)$& $-S^2 C_\xi^l$& $0$  \\ \hline
$\frac{B_T^I-B_T^{II}}{2}$&0&$S(G z_K^I- z^I)$ &$S 
(S \hat{\lambda}_{wst}^R-G \Delta \lambda_{wst}^I)$ \\ \hline
$D_T^I$&$0$ & $S(z_K^R-Gz^R)$& $-S (\Delta \lambda_{wst}^R+S\hat{\lambda}_{l}^I)  $ \\ \hline 
$ E_T^I$ &$GS$ &$0$ &$0$ \\  \hline 
$\frac{B_T^I+B_T^{II}}{R_T^I+R_T^{II}}$&$-S^2$ & 0 & 0 \\ \hline
\end{tabular}
\end{center}
\label{Tab7}
\end{table}

Although the asymmetry in eq.(\ref{T8})
is not exactly T-asymmetry,
the measurement of the coefficients are useful for constraining
$S$ and $G$ as well as various non-standard interactions. Non-standard interactions include wrong sign decay and CPT violation.  
In the following subsections, we show how one can determine $S$ and $G$
 and also show how one can constrain the various non-standard interactions.
We first study the case without any assumption and in later subsections,
we investigate two physically interesting cases, one corresponding to the case 
that CPT is a good symmetry and the other is the case without wrong sign 
decays. Since there are relations among the coefficients for different processes,
we first identify the independent coefficients.
From tables \ref{Tab3}-\ref{Tab6}, 
one finds the following relations among the coefficients for the different processes.
\bea
R_T^{IV}&=&-R^I_T,\quad  R_T^{III}=-R_T^{II}, \nn \\
C_T^{III}&=&-C_T^{II}, \quad C_T^{IV}=-C^I_T, \nn \\
S_T^{III}&=&S_T^{II}, \quad S_T^{IV}=S^I_T, \nn \\
B_T^{III}&=&-B_T^{II}, \quad B_T^{IV}=-B^I_T, \nn \\
D^I_T&=&D_T^{II}=D_T^{III}=D_T^{IV}, \nn \\
E^I_T&=&E_T^{II}=-E_T^{III}=-E_T^{IV}. \nn 
\eea
They imply that there are ten independent coefficients.
In table \ref{Tab7}, we show how ten independent combination of the coefficients can be written in terms of CPT even , CPT odd, and 
wrong sign decay parameters.
Since there are eighteen parameters, the number of the independent 
coefficients is not enough to extract the parameters.
However, one can still constrain the combination 
of the parameters.
Below we investigate how to extract the parameters for the three cases.
\subsection{Extracting the parameters from the coefficients: General case}
\label{sub41}
\begin{figure}[ht]
\centering
  \includegraphics[width=5.0cm]{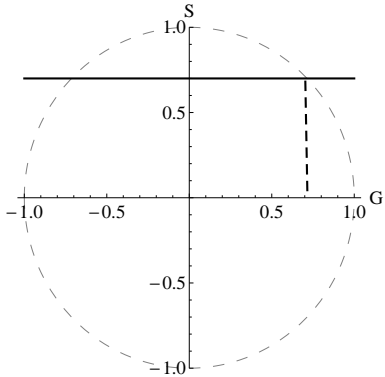}
  \caption{Determination of $G$ and $S$. Due to $G^2+S^2\simeq 1$,
  $G$ and $S$ are on the circle of unit length.
  Once $S$ is known, $G$ is determined within two-fold ambiguity.
  }
  \label{circle}
  \end{figure}

Let us first examine how one can determine the parameters from 
the measurements of the coefficients shown in table \ref{Tab7}.
Hereafter, we discuss a method to determine the values for $G$ and $S$ through observing $E_T$.
Since $E_T$ is multiplied by $y$ in eq. (\ref{T8}), one cannot
extract the value of $E_T$ solely from the measurement of
the asymmetry.
Therefore, we need to determine the value of $y$ through
the other experiment.
$y$ defined in eq. (\ref{T1p})
is regarding to the width difference of B meson mass eigenstate,
and a method to measure $y\cos \beta \simeq Gy$ is suggested in
refs. \cite{Dighe:2001gc}-\cite{Dighe:2001sr}.
Combining the measurement of $E_T^I y\simeq GSy$,
one can determine $S$.
Since $S$ and $G$ in their leading order satisfy
$S^2+G^2\simeq 1 -\mathcal{O}(C^2)$, the measurement of 
$E_T$ determines $(\pm G, S)$ within the two-fold ambiguity.
The ambiguity would be removed if we assume that the standard model
contribution is dominant for the width difference. (See figure \ref{circle})
As an alternative way to determine $S$, one can use
the relation
$\frac{B_T^I+B_T^{II}}{R_T^I+R_T^{II}}=-S^2$,
and determine $|S|$.
The sign ambiguity for $S$ can be removed because in the leading order $S$ is equal to $\frac{S_T^{II}-S_T^I}{2}$. Excluding the case that the
sub-leading contribution changes the sign of the leading term, one can determine the sign for $S$ through observing
$\frac{S_T^{II}-S_T^I}{2}$.  
Having determined $G$ and $S$, we consider
constraining the other parameters.\par
We note that the following relation is satisfied,
\bea
\frac{R_T^I-R_T^{II}}{2}+\frac{C_T^I-C_T^{II}}{2}=-\hat{\lambda}_{\mathrm{wst}}^R+G\hat{\lambda}_l^R+S \Delta \lambda_l^I.  \label{eq:DRDC}
\eea
Since the right-hand side is independent of CPT violating parameters,
non-vanishing contribution implies the unambiguous evidence of the presence of the wrong sign decay. Furthermore,
since $S$ is determined, one can write the CPT violating parameter
$\theta_K^I$ as,
\bea
\theta_K^I=\frac{\frac{S_T^I-S_T^{II}}{2}+S}{G}. \label{eq:thetakI}
\eea
If the right-hand side is non-zero, it implies
CPT violation in the right strangeness decay. However, the real part of $\theta_K$ cannot be singly extracted, since
\bea
\theta_K^R+ C=\frac{C_T^I+C_T^{II}}{2}. \label{eq:Cp}
\eea
One also notes the relation,
\bea
-R_M+\xi_l^R+C_\xi^l=\frac{R_T^I+R_T^{II}}{2}. \label{eq:Rp} 
\eea
If any one of the combinations, $\frac{R_T^I-R_T^{II}}{2},
\frac{C_T^I-C_T^{II}}{2}, \frac{S_T^I+S_T^{II}}{2},
\frac{B_T^I-B_T^{II}}{2}$, and $D_T^I$ 
is non-zero, it implies CPT violation and/or 
wrong sign decay. However, if 
the cancellation between CPT violation and wrong sign decay occurs, 
they can vanish.
\subsection{Extracting the parameters from the coefficients: CPT conserving limit}
\label{sub42}
Next we consider the case in the limit of CPT symmetry.
In the limit of CPT symmetry, all the contribution in the second column vanishes in table \ref{Tab7}. Since all the wrong sign decay parameters
are  CPT even, the third column of table \ref{Tab7} does not vanish. 
In the limit $C$, $S$ and $R_M-\xi_l^R$ can be determined as,
\bea
C&=&\frac{C_T^I+C_T^{II}}{2}, \\
S&=&\frac{S_T^{II}-S_T^I}{2},  \\
R_M-\xi_l^R&=&-\frac{R_T^I+R_T^{II}}{2}.
\eea 
Moreover T-odd wrong sign semi-leptonic decay $\Delta \lambda_l$ can be determine as, 
\bea
\Delta \lambda_l^I&=&\frac{C_T^I-C_T^{II}}{2 S},  \\
\Delta \lambda_l^R&=&-\frac{S_T^I+S_T^{II}}{2 GS}.
\eea
For the other five wrong sign decay parameters $\hat{\lambda}_{\mathrm{wst}}^R, \hat{\lambda}_l^{R,I} \Delta \lambda_{\mathrm{wst}}^{R, I}$, one obtains the following three constraints. 
\bea
\frac{R_T^I-R_T^{II}}{2}&=&-\hat{\lambda}_{\mathrm{wst}}^R+G\hat{\lambda}_l^R,
\\
\frac{B_T^I-B_T^{II}}{2}&=&S 
(S \hat{\lambda}_{\mathrm{wst}}^R-G \Delta \lambda_{\mathrm{wst}}^I), \\ 
D_T^I&=&-S (\Delta \lambda_{\mathrm{wst}}^R+S\hat{\lambda}_{l}^I). 
\eea
\subsection{Extracting the parameters from the coefficients: Case without wrong sign decay}
\label{sub43}
Lastly, we consider the case without wrong sign decays.
The relations in eqs. (\ref{eq:thetakI}-\ref{eq:Rp})  are 
satisfied in this case. The right-hand side of eq. (\ref{eq:DRDC}) vanishes.
In addition to these,
CP and CPT violation of the mixing parameters in B meson 
system is determined by
\bea
z^I=\frac{R_T^I-R_T^{II}}{2S}, \quad z^R=\frac{S_T^I+S_T^{II}}{2GS}. 
\eea   
CP and CPT violation in the neutral K meson system is also determined as,
\bea
z_K^I=\frac{D_T^I+\frac{S_T^I+S_T^{II}}{2}}{S}, \quad z_K^R=\frac{B_T^I-B_T^{II}-
(C_T^I-C_T^{II})}{2 SG}. 
\eea
The five parameters $C_\xi^l, \theta_K^R, C, R_M $ and $\xi_l^R$ 
satisfy the two constraints eqs. (\ref{eq:Cp}-\ref{eq:Rp}). 
\section{Conditions for authentic time reversed process}
\label{Sec:4}
\begin{figure}[ht]
\centering
  \includegraphics[width=7.5cm]{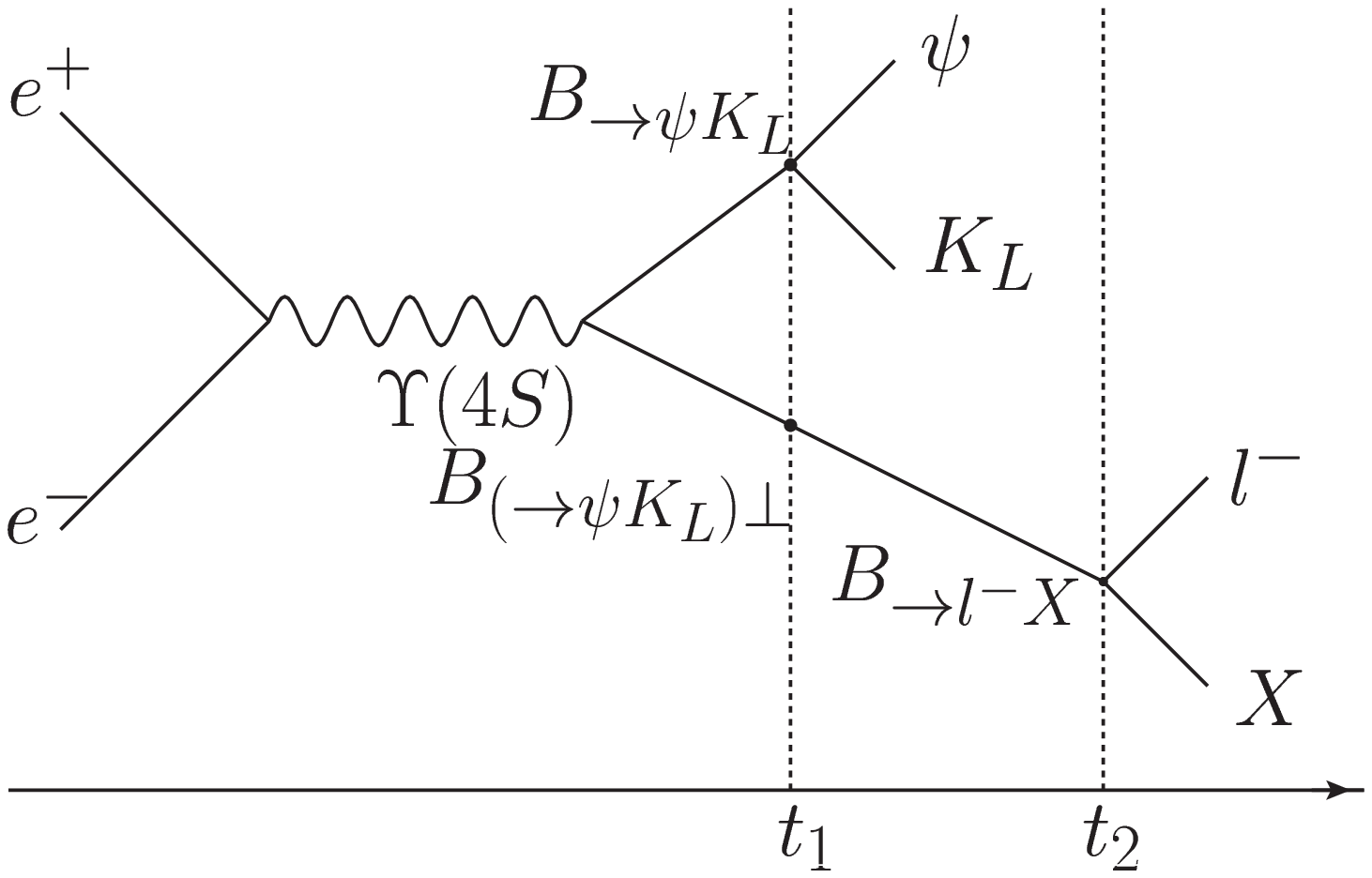}
  \caption{
  A process with $(f_1, f_2)=(\psi K_L, l^-X)$}
  \label{Fig1}
  \includegraphics[width=7.5cm]{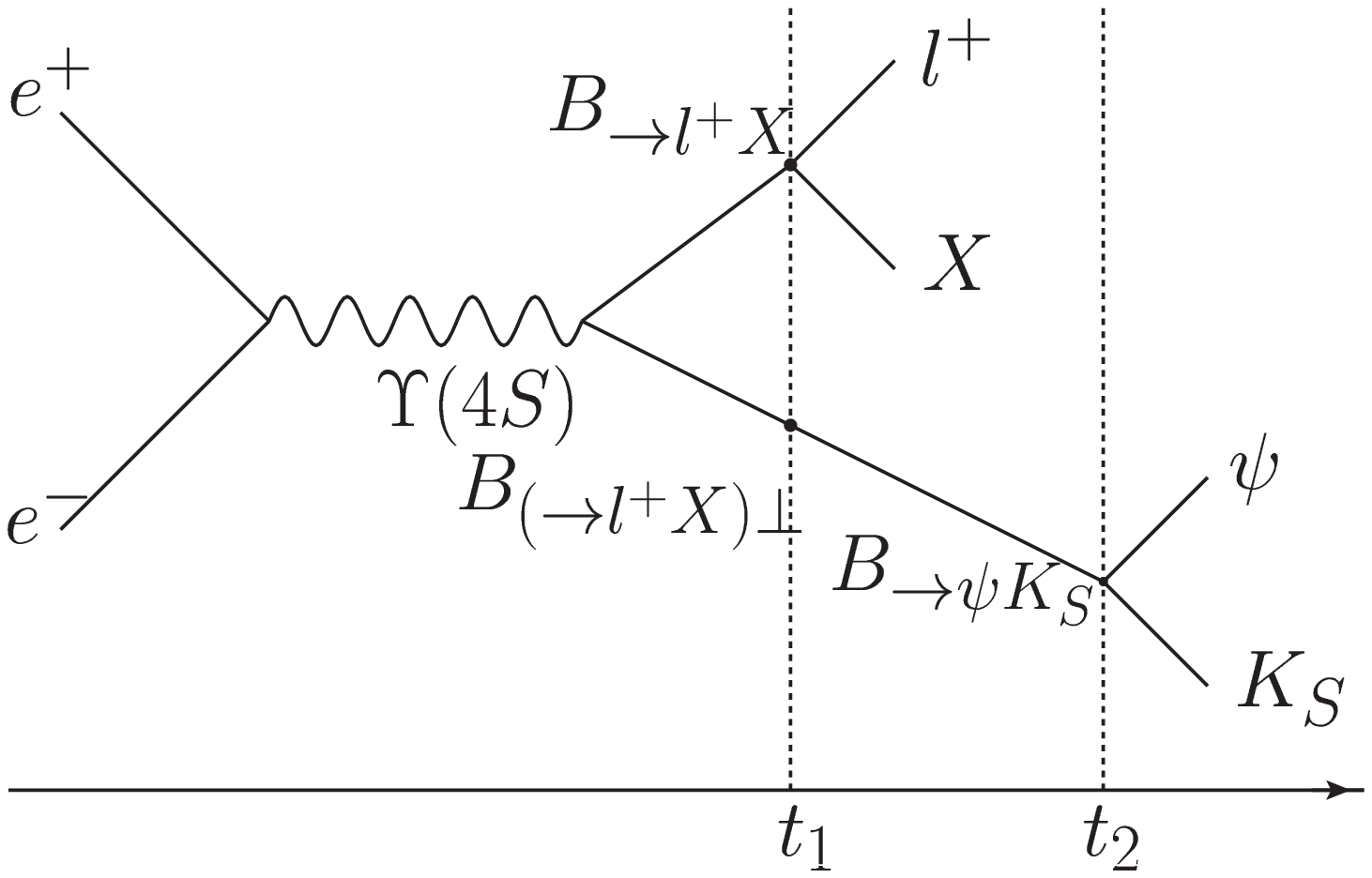}
  \caption{
  A process with $(f_3, f_4)=(l^+X, \psi K_S)$.
  figure \ref{Fig1} and figure \ref{Fig2} are referred as (I)
  in eq. (\ref{Bprocesses}).
  Event number asymmetry of figure \ref{Fig1} and figure \ref{Fig2}.
  is calculated as eqs. (\ref{T9})-(\ref{T14})
  }
  \label{Fig2}
  \includegraphics[width=7.5cm]{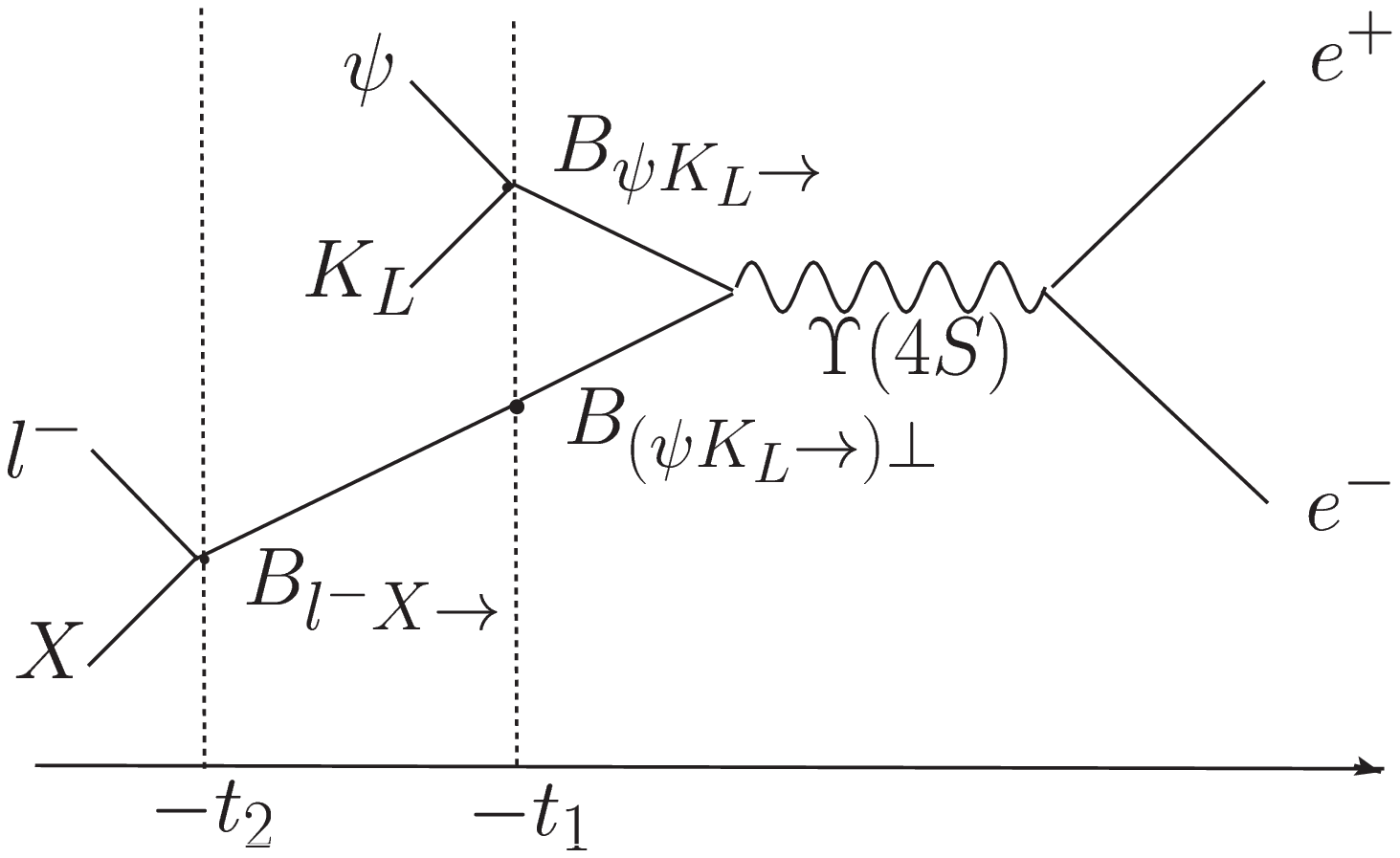}
  \caption{A process with inverse decays of B meson.
  figure \ref{Fig1} and figure \ref{Fig3} are related with flipping time
  direction.}
  \label{Fig3}
  \includegraphics[width=7.5cm]{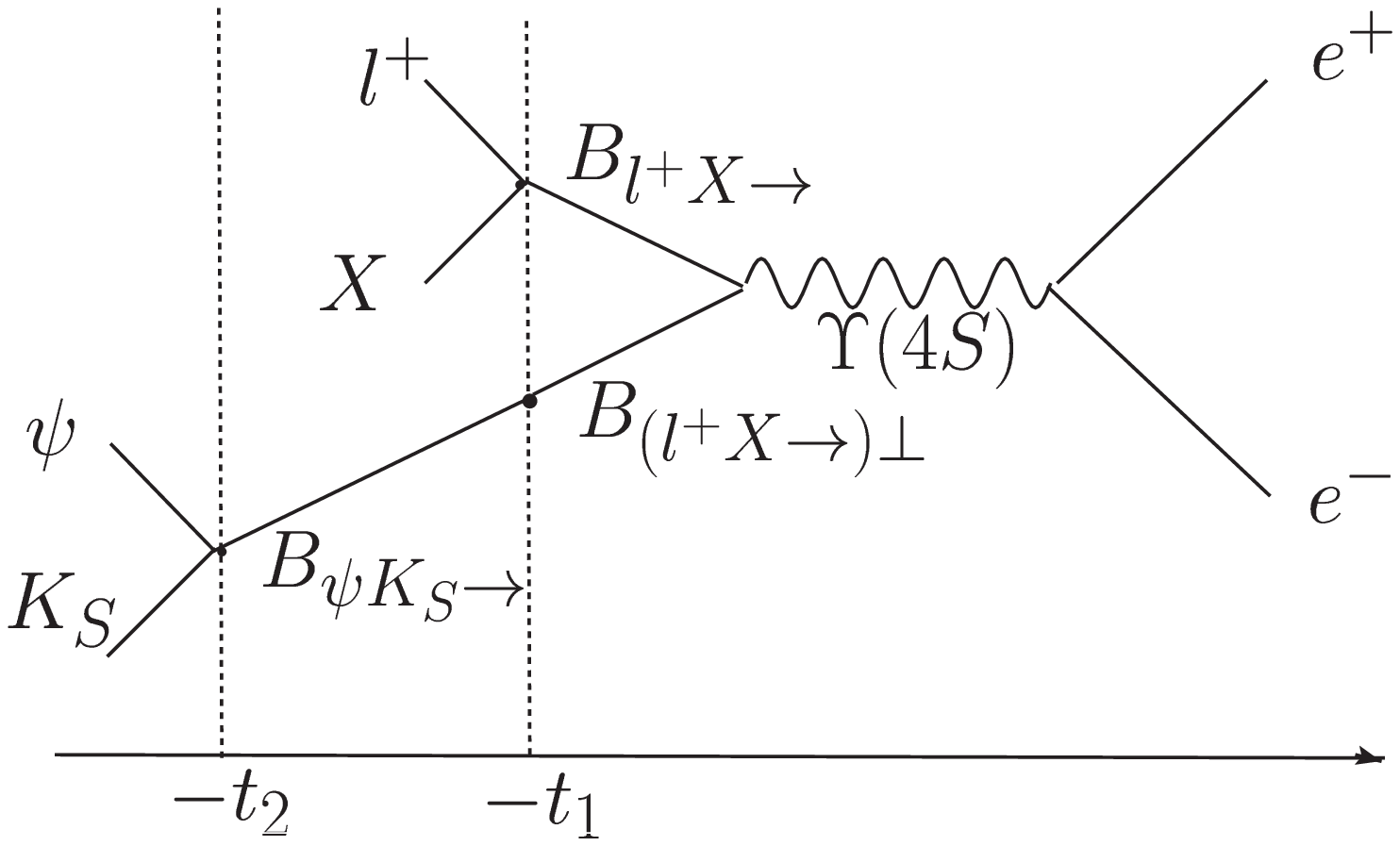}
  \caption{A process with inverse decays of B meson.
  figure \ref{Fig2} and figure \ref{Fig4} are related with flipping time
  direction.}
  \label{Fig4}
\end{figure}

In section \ref{Sec:2}, we showed the expression of 
the asymmetry that describes event number difference of
figure \ref{Fig1} and figure \ref{Fig2}.
However, rather than figure \ref{Fig2},
figure \ref{Fig3} is an authentic time reversed
process of figure \ref{Fig1},
since the two processes of figure \ref{Fig1} and figure \ref{Fig3} 
are related with flipping time direction.
In discussion given in refs.\cite{Banuls:1999aj}-\cite{Bernabeu:2012ab},
one substituted figure \ref{Fig2} for figure \ref{Fig3}
because signal sides of
figure \ref{Fig1} and figure \ref{Fig2}
are deemed to be a time reversed process to each other.
Since figure \ref{Fig2} is not an authentic time reversed process,
the asymmetry is slightly deviated from T-odd.\par
In this section, we clarify why T-even parts are included in the coefficients
eqs.(\ref{T9}-\ref{T13}), 
although it is naively thought to be a T-odd quantity.
\\
One can show that, when the following conditions are simultaneously satisfied, 
figure \ref{Fig2} plays the role as a time reversed process of figure \ref{Fig1}
and the coefficients eqs.(\ref{T9}-\ref{T13}) become T-odd.
\begin{itemize}
\item[1.] Equivalence conditions of B meson states. 
\item[2.] $\Delta N_R^{\mathrm{e}}=0$. 
\end{itemize}
where we denote $\Delta N_R=\Delta N_R^{\mathrm{o}}+\Delta N_R^{\mathrm{e}}$
and $\Delta N_R^\mathrm{e}$ $(\Delta N_R^\mathrm{o})$ is the T-even (odd) part.\par
The equivalence conditions indicate that
the initial (final) B meson states of signal side in figures.\ref{Fig2}-\ref{Fig3}
are the same as each other.
The equivalence conditions are described as,
\bea
  \left\{
    \begin{array}{l}
      \ket{ B_{(\rightarrow l^+X)\perp}}\propto \ket{B_{l^-X\rightarrow}}\\
      \ket{ B_{\rightarrow \psi K_S}}\propto \ket{B_{(\psi K_L\rightarrow)\perp}}
    \end{array} \right..
\label{J1}
\eea
 Eq. (\ref{J1}) shows that B meson states in figures \ref{Fig2}-\ref{Fig3} are equivalent.
Similarly, figure \ref{Fig4} is the authentic time reversed process
of figure \ref{Fig2}.
When we apply the same condition to
B meson states in figures \ref{Fig1}-\ref{Fig4}, one obtains,
\bea
  \left\{
    \begin{array}{l}
\ket{ B_{(\rightarrow l^- X)\perp}}\propto \ket{B_{(l^+ X\rightarrow)}}\\
      \ket{ B_{(\rightarrow \psi K_L)}}\propto \ket{B_{(\psi 
K_S\rightarrow})_\perp}
    \end{array} \right..
\label{J3}
\eea
Violation of the conditions (\ref{J1}-\ref{J3}) is originally calculated in ref. \cite{Applebaum:2013wxa}.
Including overall factors and using our notation, we show the violation of the conditions(\ref{J1}-\ref{J3}) as follows,
\bea
  \left\{
    \begin{array}{l}

      \braket{B_{(\psi K_L\rightarrow)\perp}|B_{(\rightarrow \psi K_S)\perp}}=\displaystyle N_{(\rightarrow \psi K_S)\perp}N_{(\psi K_L \rightarrow)\perp}(A_{\psi K^0}A_{\psi K^0}^{\mathrm{ID}}+\bar{A}_{\psi \bar{K^0}}\bar{A}_{\psi \bar{K^0}}^{\mathrm{ID}})
\frac{\theta_K+\Delta\lambda_{\mathrm{wst}}}{2}\\
\braket{B_{(l^-X\rightarrow)\perp}|B_{(\rightarrow l^+X)\perp}}=2N_{(l^-\rightarrow )\perp}N_{(\rightarrow l^+)\perp}A_{l^+}\bar A_{l^-}^{\mathrm{ID}}\displaystyle\frac{p}{q}\lambda_{l^+},
    \end{array}
  \right.
\label{J2}\\
\left\{
    \begin{array}{l}
      \braket{B_{(\psi K_S\rightarrow)\perp}|B_{(\rightarrow \psi K_L)\perp}}=\displaystyle N_{(\rightarrow \psi K_L)\perp}N_{(\psi K_S \rightarrow)\perp}(A_{\psi K^0}A_{\psi K^0}^{\mathrm{ID}}+\bar{A}_{\psi \bar{K^0}}\bar{A}_{\psi \bar{K^0}}^{\mathrm{ID}})
\frac{\theta_K-\Delta\lambda_{\mathrm{wst}}}{2}\\
\braket{B_{(l^+X\rightarrow)\perp}|B_{(\rightarrow l^-X)\perp}}=2N_{(l^+\rightarrow )\perp}N_{(\rightarrow l^-)\perp}\bar A_{l^-}A_{l^+}^{\mathrm{ID}}\displaystyle\frac{q}{p}\lambda_{l^-}^{-1},
    \end{array}
  \right.
\label{J4}
\eea
where we used the expression for states defined in
eqs. (\ref{Bket}-\ref{EE8}, \ref{EE5}-\ref{EE10}).
In eqs. (\ref{J2}-\ref{J4}), effect of mixing-induced CP violation in K meson
system is included in terms of our notation $\Delta\lambda_{\mathrm{wst}}
\simeq\displaystyle\frac{p_K}{q_K}\displaystyle\frac{A_{\psi \bar{K^0}}}{A_{\psi K^0}}-\displaystyle\frac{q_K}{p_K}\displaystyle\frac{\bar{A}_{\psi K^0}}{\bar{A}_{\psi \bar{K^0}}}$ in comparison with ref. \cite{Applebaum:2013wxa}.\\
$\braket{B_{(l^-X\rightarrow)\perp}|B_{(\rightarrow l^+X)\perp}}\ne 0$ and $\braket{B_{(l^+X\rightarrow)\perp}|B_{(\rightarrow l^-X)\perp}}\ne0$ indicate that one cannot exactly conduct the flavor tagging in the presence of wrong sign semi-leptonic decays.
Similarly,
$\braket{B_{(\rightarrow\psi K_S)\perp}|B_{(\psi K_L\rightarrow)\perp}}\ne0$
and $\braket{B_{(\psi K_S\rightarrow)\perp}|B_{(\rightarrow \psi K_L)\perp}}\ne0$
imply that one cannot exactly carry out the CP tagging
in the presence of CPT violation in decays and wrong sign strangeness decays.
Therefore, eqs. (\ref{J2}-\ref{J4}) describe that
semi-leptonic decays
and strangeness changing decays yield
tagging ambiguities, and that are expressed in terms of
state non-orthogonality.
\par
Then, we turn to explanation of the second condition, $\Delta N_R^\mathrm{e}=0$.
We define the following quantities for the expedient sake.
\bea
X^{\mathrm{o}}=\frac{X_{(\psi K_L)\perp, l^+X}}{\kappa_{(\psi K_L)\perp, l^+X}}-
\frac{X_{(l^-X)\perp, \psi K_S}}{\kappa_{(l^-X)\perp, \psi K_S}},\quad
X^{\mathrm{e}}=\frac{X_{(\psi K_L)\perp, l^+X}}{\kappa_{(\psi K_L)\perp, l^+X}}+
\frac{X_{(l^-X)\perp, \psi K_S}}{\kappa_{(l^-X)\perp, \psi K_S}},
\label{NR1}
\eea
where $X=\sigma, \mathcal{C}$ and $\mathcal{S}$ are given
in eqs. (\ref{M3}-\ref{M5}).
Consider the case that the equivalence conditions are satisfied 
to demonstrate that violation of $\Delta N_R^\mathrm{e}=0$ gives rise to T-even
contribution to the asymmetry.
For that case, we can find that $X^\mathrm{o} (X^\mathrm{e})$ defined in eq. (\ref{NR1})
is T-odd (even) due to expressions given as follows,
\bea
&S^{\mathrm{o}}=-2S(1-Gz^R),\quad
C^{\mathrm{o}}=2[C-Sz^I],\quad
(\sigma^{\mathrm{o}})^l= 0,\label{So}&\\
&S^{\mathrm{e}}=2[Gz_K^I+(S^2-1)z^I]\label{Se},\quad
C^{\mathrm{e}}=2[z_K^R-Gz^R],\quad
(\sigma^{\mathrm{e}})^l= 2G,&
\eea
where for $\Delta\sigma$ and $\hat{\sigma}$, we write down only the 
leading part since small parts of $\Delta\sigma$ and $\hat{\sigma}$
are neglected when multiplied by $y$ in the asymmetry (\ref{T6}).
For the process (I), $\Delta X$ and $\hat{X}$ defined in eqs. (\ref{T4}-\ref{T5}) are
written as,
\bea
\Delta X \simeq X^{\mathrm{o}}-\frac{\Delta N_R}{2} X^{\mathrm{e}}=\left(X^\mathrm{o}-\frac{\Delta N_R^\mathrm{o}}{2}X^\mathrm{e}\right)-\frac{\Delta N_R^\mathrm{e}}{2}X^\mathrm{e}
,
\label{delX}\\
\hat{X} \simeq X^{\mathrm{e}}-\frac{\Delta N_R}{2} X^{\mathrm{o}}
=\left(X^\mathrm{e}-\frac{\Delta N_R^\mathrm{o}}{2}X^\mathrm{o}\right)-\frac{\Delta N_R^\mathrm{e}}{2}X^\mathrm{o}
.\label{hatX}
\eea
One finds that the T-even part of $\Delta N_R$ leads T-even contribution to $\Delta X$ in eq. (\ref{delX}).
The same applies to $\hat{X}$,
and it is shown that $\Delta X$ $(\hat{X})$ deviates from T-odd (even)
when $\Delta N_R^\mathrm{e}$ has non-zero value.
Therefore, we can demonstrate that $\Delta N_R^\mathrm{e}$
gives rise to T-even contribution to the asymmetry given in
eqs. (\ref{RT}-\ref{ET}).
\section{Conclusion}
\label{Conclusion}
In this paper, the precise meaning of the time reversal-like 
asymmetry is investigated, based on the most general
time dependence of the asymmetry in eq.(\ref{T8}).
In analysis of BaBar \cite{Lees:2012uka} and \cite{Applebaum:2013wxa},
the difference of the overall constants for the rates is eliminated.
The ratio of the overall constants for the two decay rates is deviated from unity,
and the deviation $\Delta N_R=N_R-1$
is taken into account in our analysis.
If one takes
the limits $\Delta N_R=0$ and $y=0$ in our analysis,
the asymmetry of BaBar collaboration \cite{Lees:2012uka} is obtained.
In our analysis, since the final states $\psi K_{S, L}$ are not the exact CP eigenstates, one can find
the effect of mixing-induced CP violation in K meson system.
The effect of $\epsilon_K$ is extracted and it gives rise to
$\mathcal{O}(10^{-3})$ contribution to $C_T$,
the coefficient of $\cos(x\Gamma t)$.
$\epsilon_K^R$ and direct CP violation
$|\bar{A}_{\psi \bar{K^0}}/A_{\psi K^0}|$
are dominant in $C_T$, if the wrong sign semi-leptonic
decay and CPT violations are negligible.
As well as $\epsilon_K$, the contribution from CPT violation in Kaon system $z_K$
is estimated.
\par
We introduced the parameters which have the specific property under CP, T and CPT transformations,
including the effect of indirect CP violation in K meson
system.
Taking account of the difference for overall constants,
the coefficients of each time-dependent function
are written in terms of such parameters,
and one can find that the asymmetry
consists of not only T-odd terms, but also T-even terms
in the most general time dependent function for the
asymmetry.
Furthermore, the introduced parameters are invariant
under rephasing of quarks.
We also found that the asymmetry is expressed
as phase convention independent quantities.
\par
We obtained the coefficients of the asymmetry for
the processes (I-IV) and
studied how to extract the parameters.
Assuming that the value of $y$, i.e., the width difference of $B_d$
meson mass eigenstates is known,
the three cases to constrain the parameters are discussed.
For the most general case,
combining the coefficients for different processes enables one to determine the parameters, $S$ and $G$.
We also find that non-zero value of some combination of the coefficient signals either CPT violation or the presence of the wrong sign decays.
The other two cases correspond to CPT-conserving limit and the absence of wrong sign 
decays.
In the CPT-conserving case, the coefficients constrain the parameters for
wrong sign decays.
In the absence of wrong sign decays, indirect CPT violation for B meson and K meson is constrained.\par
Moreover, we discussed T-even parts in the asymmetry.
We found that T-even terms in the asymmetry
vanish when several conditions are satisfied.
These derived conditions are categorized as two parts.
The first one is referred as equivalence conditions,
regarding to B meson states for a time reversal-like process and an authentic time reversed process.
As suggested in \cite{Applebaum:2013wxa},
B mesons for the two processes are not equivalent
to each other, and we also showed the violation of the equivalence conditions, including the effects of
mixing in K meson system.
Since non-zero $\Delta N_R$ is taken into account
in our study, $\Delta N_R$ can be the origin of T-even contribution to
the asymmetry.
We investigated that the second condition, which requires that T-even parts of $\Delta N_R$ are zero, is needed for the asymmetry
to become a T-odd quantity.
\appendix
\section{Coefficients of master formula}
\label{Sec:C}
We record coefficients of the master formula for the time-dependent decay rate of ref. \cite{Applebaum:2013wxa}.
  \bea
    N_{(i)\perp, j}&=&\frac{1}{4} {\cal N}_i {\cal N}_j \{ 1+ (C_i+C_j) (R_M-z^R) \},\quad
{\cal N}_i=|A_i|^2+|\bar{A}_i|^2,
    \label{M1} \\
    \kappa_{(i) \perp,j}&=&(1-G_i G_j) \nn \\
      &&+ [(C_i+C_j)(1-G_i G_j)+ C_j G_i+C_i G_j] z^R-(S_i+S_j) z^I \nn \\
      && +G_i G_j (C_i \theta^R_i+C_j \theta^R_j) -G_i S_j \theta^I_j-G_j S_i \theta^I_i, \label{M2}\\
    \sigma_{(i)\perp, j}&=&G_j-G_i \nn \\
      &&+[C_i(1+G_j-G_i)-C_j(1-G_j+G_i)]z^R+(G_iS_j-G_jS_i)z^I \nn \\
      &&-C_jG_j\theta^R_j+S_j\theta^I_j+C_iG_i\theta^R_i-S_i\theta^I_i,\label{M3} \\
    \C_{(i)\perp, j}&=&-C_iC_j-S_iS_j \nn \\
      &&-[(C_i+C_j)(C_iC_j+S_iS_j)+C_iG_j+C_jG_i]z^R+(S_i+S_j)z^I\nn \\
      &&+G_jS_i\theta^I_j-[C_i(1-C_j^2)-C_jS_iS_j]\theta^R_j\nn \\
      &&+G_iS_j\theta^I_i-[C_j(1-C_i^2)-C_iS_iS_j]\theta^R_i,\label{M4} \\
    \S_{(i)\perp, j}&=&C_iS_j-C_jS_i\nn \\
      &&+[C_iC_j(S_j-S_i)-(C_j^2+G_j)S_i+(C_i^2+G_i)S_j]z^R+(C_j-C_i)z^I\nn \\
      &&-C_iG_j\theta^I_j+[(C_j^2-1)S_i-C_iC_jS_j]\theta^R_j\nn \\
      &&+C_jG_i\theta^I_i-[(C_i^2-1)S_j-C_iC_jS_i]\theta^R_i, \label{M5}
  \eea
where $A_i$ and $\bar{A}_i$ in eq.(\ref{M1})
are defined in eq.(\ref{K2}).
$i$ and $j$ represent the final state of tagging side ($f_i$) and signal side($f_j$) for a pair of B meson decaying respectively.
\section{Incoming mass eigenstates and outgoing mass eigenstates in B meson and K meson system}
\label{Sec:D}
Throughout this paper, 
the time reversal process of B meson decay often appears.
To describe the inverse decay amplitudes, as out-states of B mesons,
the reciprocal
base must be used for Non-Hermitian Hamiltonian system.
This is formulated in several literatures,
refs. \cite{Sachs:1963zz}-\cite{Silva:2000db}.
In this appendix,
we show the definition of incoming states and outgoing states which are 
used in this paper.

The incoming mass eigenstates of effective Hamiltonian in B meson system are
\bea
 \ket{B_H^{in}}&=&p\sqrt{1+z}\ket{B^0}-q\sqrt{1-z}\ket{\bar B^0}\label{BH},\\
 \ket{B_L^{in}}&=&p\sqrt{1-z}\ket{B^0}+q\sqrt{1+z}\ket{\bar B^0}\label{BL},
\eea
where $p$, $q$ are mixing parameters in indirect CP violation and $z$ is a mixing parameter in indirect CP, CPT violation.
In terms of matrix elements of effective Hamiltonian, these parameters are written as 
\bea
 \frac{p}{q}&=&\sqrt{\frac{M_{12}-\frac{i}{2}\Gamma_{12}}{M_{12}^*-\frac{i}{2}\Gamma_{12}^*}},\label{poverq}\\
 z&=&\frac{M_{11}-M_{22}-\frac{i}{2}(\Gamma_{11}-\Gamma_{22})}{m_{L}-m_{H}-\frac{i}{2}(\Gamma_{L}-\Gamma_{H})}\label{zet}.
\eea
These expressions help us understand the transformation property of $\lambda$, $S$, $G$, $C$, etc.
Then, outgoing mass states are determined to fulfill the following conditions,
\bea
 \braket{B_H^{out}|B_H^{in}}=1,\braket{B_H^{out}|B_L^{in}}=0,\braket{B_L^{out}|B_L^{in}}=1,\braket{B_L^{out}|B_H^{in}}=0.
\eea
Thus, outgoing mass eigenstates are
\bea
 \bra{B_H^{out}}&=&\frac{1}{2pq}(q\sqrt{1+z}\bra{B^0}-p\sqrt{1-z}\bra{\bar B^0}) ,\\
 \bra{B_L^{out}}&=&\frac{1}{2pq}(q\sqrt{1-z}\bra{B^0}+p\sqrt{1+z}\bra{\bar B^0}).
\eea
For K meson system, similarly,
\bea
 \ket{K_L^{in}}&=&p_K\sqrt{1+z_K}\ket{K^0}-q_K\sqrt{1-z_K}\ket{\bar K^0}
 ,\label{KLin}\\
 \ket{K_S^{in}}&=&p_K\sqrt{1-z_K}\ket{K^0}+q_K\sqrt{1+z_K}\ket{\bar K^0},
\label{KSin}\\
 \bra{K_L^{out}}&=&\frac{1}{2p_Kq_K}(q_K\sqrt{1+z_K}\bra{K^0}-p_K\sqrt{1-z_K}\bra{\bar K^0}), \label{KLout}\\
 \bra{K_S^{out}}&=&\frac{1}{2p_Kq_K}(q_K\sqrt{1-z_K}\bra{K^0}+p_K\sqrt{1+z_K}\bra{\bar K^0}).\label{KSout}
\eea
Since these mass eigenstates in K meson system are shown in flavor states,
we specifically can calculate the amplitudes of transition from incoming states B meson to outgoing states $\psi K_{L, S}$
\bea
 A_{\psi K_S}&=& \braket{\psi K_S^{out}|B_0^{in}}
                        =\frac{1}{2p_Kq_K}(q_K\sqrt{1-z_K}A_{\psi K^0}+p_K\sqrt{1+z_K}A_{\psi \bar{K^0}}),\label{D1}\\
 A_{\psi K_L}&=& \braket{\psi K_L^{out}|B_0^{in}}
                        =\frac{1}{2p_Kq_K}(q_K\sqrt{1+z_K}A_{\psi K^0}-p_K\sqrt{1-z_K}A_{\psi \bar{K^0}}),\label{D2}\\
 \bar{A}_{\psi K_S}&=& \braket{\psi K_S^{out}|\bar B_0^{in}}
                                =\frac{1}{2p_Kq_K}(q_K\sqrt{1-z_K}\bar{A}_{\psi K^0}+p_K\sqrt{1+z_K}\bar{A}_{\psi \bar{K^0}}),\label{D3}\\
 \bar{A}_{\psi K_L}&=& \braket{\psi K_L^{out}|\bar B_0^{in}}
                                =\frac{1}{2p_Kq_K}(q_K\sqrt{1+z_K}\bar{A}_{\psi K^0}-p_K\sqrt{1-z_K}\bar{A}_{\psi \bar{K^0}}).\label{D4}
\eea
Hence, we can obtain eqs. (\ref{K7}), (\ref{K8}), (\ref{K9}) and (\ref{K4}).
We also can write down inverse decay amplitudes of eqs.
(\ref{D1}-\ref{D4})
\bea
 A_{\psi K_S}^{\mathrm{ID}}&=& \braket{B_0^{out}|\psi K_S^{in}}
                        =(p_K\sqrt{1-z_K}A_{\psi K^0}^{\mathrm{ID}}+q_K\sqrt{1+z_K}A_{\psi \bar{K^0}}^{\mathrm{ID}}),\label{D5}\\
 A_{\psi K_L}^{\mathrm{ID}}&=& \braket{B_0^{out}|\psi K_L^{in}}
                        =(p_K\sqrt{1+z_K}A_{\psi K^0}^{\mathrm{ID}}-q_K\sqrt{1-z_K}A_{\psi \bar{K^0}}^{\mathrm{ID}}),\label{D6}\\
 \bar{A}_{\psi K_S}^{\mathrm{ID}}&=& \braket{\bar B_0^{out}|\psi K_S^{in}}
                                =(p_K\sqrt{1-z_K}\bar{A}_{\psi K^0}^{\mathrm{ID}}+q_K\sqrt{1+z_K}\bar{A}_{\psi \bar{K^0}}^{\mathrm{ID}}),\label{D7}\\
 \bar{A}_{\psi K_L}^{\mathrm{ID}}&=& \braket{\bar B_0^{out}|\psi K_L^{in}}
                                =(p_K\sqrt{1+z_K}\bar{A}_{\psi K^0}^{\mathrm{ID}}-q_K\sqrt{1-z_K}\bar{A}_{\psi \bar{K^0}}^{\mathrm{ID}}).\label{D8}
\eea

\section{List of coefficients of time dependent decay rates for process (I)}
\label{Sec:A}
In this appendix, we show the coefficients of the time dependent decay rates
in eq. (\ref{T1}) which are needed for calculation of the asymmetry
of process (I).
\bea
S_{(\psi K_L)\perp, l^-X}&=&\SKL-\SKL z^R-z^I-\GKL\theta_{\psi K_L}^I,\label{appC1}\\
S_{(l^+X)\perp, \psi K_S}&=&\SKS+\SKS z^R-z^I-\GKS\theta_{\psi K_S}^I,\\
C_{(\psi K_L)\perp, l^-X}&=&\CKL-\SKL S_{l^-}+\GKL z^R+\SKL z^I+\theta_{\psi K_L}^R,\\
C_{(l^+X)\perp, \psi K_S}&=&-\CKS-\SKS S_{l^+}-\GKS z^R+\SKS z^I-\theta_{\psi K_S}^R, \\
\kappa_{(\psi K_L)\perp, l^-X}&=&1-\GKL G_{l^-}-(\GKL+1)z^R
-\SKL z^I, \label{eq:kappaLm} \\
\kappa_{(l^+X)\perp, \psi K_S}&=&1-\GKS G_{l^+}+(\GKS+1)z^R
-\SKS z^I, \label{eq:kappapS} \\
\sigma_{(\psi K_L)\perp, l^-X}&=&G_{l^-}-G_{\psi K_L}
+(1+G_{\psi K_L}) z^R-S_{\psi K_L} \theta^I_{\psi K_L},
\\
\sigma_{(l^+X)\perp \psi K_S}&=& G_{\psi K_S} -G_{l^+}+(1+G_{\psi K_S}) z^R
+S_{\psi K_S} \theta^I_{\psi K_S},\\
\frac{S_{(\psi K_L)\perp, l^-X}}{\kappa_{(\psi K_L)\perp, l^-X}}&=&
\SKL+\SKL \GKL G_{l^-}+\SKL\GKL z^R+(\SKL^2-1)z^I-\GKL
\theta_{\psi K_L}^I,\nn\\
\\
\frac{S_{(l^+X)\perp, \psi K_S}}{\kappa_{(l^+X)\perp, \psi K_S}}&=&
\SKS+\SKS \GKS G_{l^+}-\SKS\GKS z^R+(\SKS^2-1)z^I-\GKS
\theta_{\psi K_S}^I,\nn\\
\\
\frac{C_{(\psi K_L)\perp, l^-X}}{\kappa_{(\psi K_L)\perp, l^-X}}&\simeq&
C_{(\psi K_L)\perp, l^-X},\\
\frac{C_{(l^+X)\perp, \psi K_S}}{\kappa_{(l^+X)\perp, \psi K_S}}&\simeq&
C_{(l^+X)\perp, \psi K_S}, \\
\frac{\sigma_{(\psi K_L)\perp, l^-X}}{\kappa_{(\psi K_L)\perp, l^-X}}
&\simeq& -G_{\psi K_L},\\
\frac{\sigma_{(l^+X)\perp, \psi K_S}}{\kappa_{(l^+X)\perp, \psi K_S}}
&\simeq& G_{\psi K_S}\label{appC14},
\eea
where we keep only the leading term for $\frac{\sigma}{\kappa}$, since it will be multiplied by a small number $y$ in the formulae of the decay rate
eq. (\ref{T1}). 
\section{Expressions for $N_R, \Delta \S, \Delta \C, \Delta \sigma,
\hat{\sigma}, \hat{\S}$ and $\hat{\C}$}
\label{Sec:B}
The quantity $N_R$, defined in eq.(\ref{T3}), denotes the ratio of a normalization
for rates.
Since we compute the asymmetry including the effect
of $N_R$, its expression should be clarified.
In this appendix section, we calculate $N_R$, and obtain the expressions of
parameters as $\Delta \S, \Delta \C, \Delta \sigma,
\hat{\sigma}, \hat{\S}$ and $\hat{\C}$ for the process (I).
In deriving formulae, we use eqs. (\ref{appC1})-(\ref{appC14}).
Expanding $N_R$ with respect to small parameters, we obtain
the general structure of $N_R$ at first order approximation.
\bea
N_R&=&\frac{N_{(3)\perp, 4}}{N_{(1)\perp, 2}}\frac{\kappa_{(3)\perp, 4}}
{\kappa_{(1)\perp, 2}}\nn\\
&=&\frac{\mathcal{N}_3\mathcal{N}_4[1+(C_3+C_4)(R_M-z^R)]
}{\mathcal{N}_1\mathcal{N}_2[1+(C_1+C_2)(R_M-z^R)]}
\frac{\kappa_{(3)\perp, 4}^{l}\left(1+\displaystyle\frac{\Delta\kappa_{(3)\perp, 4}}{\kappa_{(3)\perp, 4}^{l}}\right)}
{\kappa_{(1)\perp, 2}^{l}\left(1+\displaystyle\frac{\Delta\kappa_{(1)\perp, 2}}{\kappa_{(1)\perp, 2}^{l}}\right)}\nn\\
&\simeq&
\frac{\mathcal{N}_3\mathcal{N}_4}{\mathcal{N}_1\mathcal{N}_2}
\frac{\kappa_{(3)\perp, 4}^l}{\kappa_{(1)\perp, 2}^l}
\left[
1+(C_3+C_4-C_1-C_2)(R_M-z^R)+\displaystyle\frac{\Delta\kappa_{(3)\perp, 4}}{\kappa_{(3)\perp, 4}^{l}}-\displaystyle\frac{\Delta\kappa_{(1)\perp, 2}}{\kappa_{(1)\perp, 2}^{l}}
\right], \nn\\
\label{A1}
\eea
where superscript $^l$ expresses the leading part and $\Delta$ expresses
the small part such as,
$\kappa_{(1)\perp, 2}=\kappa_{(1)\perp, 2}^l+\Delta\kappa_{(1)\perp, 2}$ and $\kappa_{(3)\perp, 4}=\kappa_{(3)\perp, 4}^l+\Delta\kappa_{(3)\perp, 4}$.
For the processes given in eq.(\ref{Bprocesses}),
$\kappa_{(1)\perp, 2}^l=\kappa_{(3)\perp, 4}^l=1$
($f_1=\psi K_L, f_2=l^-X, f_3=l^+X, f_4=\psi K_S$ for process (I))
is satisfied and $N^{I}_R$ is written as,
\bea
N^{I}_R&=&
\frac{\mathcal{N}_{l^+X}\mathcal{N}_{\psi K_S}}
{\mathcal{N}_{\psi K_L}\mathcal{N}_{l^-X}}
\left[1+(C_{\psi K_S}-C_{\psi K_L}+C_{l^+}-C_{l^-})(R_M-z^R) \right. \nn \\
&+& \left. 
\Delta \kappa_{(l^+X)\perp, \psi K_S}-\Delta \kappa_{(\psi K_L)\perp, l^-X}
\right]\nn\\
&\simeq&
\frac{\mathcal{N}_{\psi K_S}\mathcal{N}_{l^+X}}{\mathcal{N}_{\psi K_L}\mathcal{N}_{l^-X}}
[1+2(R_M-z^R)
+2(z^R-Sz^I-G\hat{\lambda}_l^R)]\nn\\
&=&
\frac{\mathcal{N}_{\psi K_S}\mathcal{N}_{l^+X}}{\mathcal{N}_{\psi K_L}\mathcal{N}_{l^-X}}
[1+
2(-Sz^I+R_M-G\hat{\lambda}_l^R)].
\label{A2}
\eea 
Deviation of $\mathcal{N}_{l^+X}/\mathcal{N}_{l^-X}$ and
$\mathcal{N}_{\psi K_S}/\mathcal{N}_{\psi K_L}$ from 1 is written in terms of small parameters as,
\bea
\frac{\mathcal{N}_{l^+X}}{\mathcal{N}_{l^-X}}=1-2(C_{\xi}^l+\xi^R_l),\quad
\label{A3}
\frac{\mathcal{N}_{\psi K_S}}{\mathcal{N}_{\psi K_L}}=1+2\hat{\lambda}_{\mathrm{wst}}^R.
\eea
\bea
N^{I}_R=1+\Delta N^{I}_R=1+2[-Sz^I+R_M+\hat{\lambda}_{\mathrm{wst}}^R-G
\hat{\lambda}_l^{R}-C^l_\xi-\xi_l^R]. \label{NR}
\eea
Note that $\Delta N^{I}_R$ is a small number.\\
We can also write down the expressions of $\Delta S^I$ and $\Delta C^I$.
\bea
\Delta \S^I&=
&\left(\frac{\S_{(\psi K_L)\perp, l^-X}}{\kappa_{(\psi K_L)\perp, l^-X}}
-\frac{\S_{(l^+X)\perp, \psi K_S}}{\kappa_{(l^+X)\perp, \psi K_S}}\right)
-\frac{\Delta N^{I}_R}{2}
\left(\frac{\S_{(\psi K_L)\perp, l^-X}}{\kappa_{(\psi K_L)\perp, l^-X}}+
\frac{\S_{(l^+X)\perp, \psi K_S}}{\kappa_{(l^+X)\perp, \psi K_S}}
\right)
\nn\\
&\simeq&
\frac{\S_{(\psi K_L)\perp, l^-X}}{\kappa_{(\psi K_L)\perp, l^-X}}
-\frac{\S_{(l^+X)\perp, \psi K_S}}{\kappa_{(l^+X)\perp, \psi K_S}}
=
-2[S(1-Gz^R)-G\theta_K^I+GS\Delta\lambda_l^R], \label{delS}
\\
\Delta \C^I&\simeq
&\frac{\C_{(\psi K_L)\perp, l^-X}}{\kappa_{(\psi K_L)\perp, l^-X}}
-\frac{\C_{(l^+X)\perp, \psi K_S}}{\kappa_{(l^+X)\perp, \psi K_S}}
=2[C-Sz^I +\theta_K^R+ S\Delta\lambda_l^I]. \label{delC}
\eea
We calculate only the leading part of $\Delta\sigma^I$ and $\hat{\sigma}^I$,
since the sub-leading part of $\Delta\sigma^I$ and $\hat{\sigma}^I$
is suppressed when multiplied with $y\Gamma t$.
\bea
\Delta \sigma^{Il}=0, \quad \hat{\sigma}^{Il}=2G.
\label{A10}
\eea
We write down the expressions for $\hat{\S}^I$ and $\hat{\C}^I$ as follows,
\bea
\hat{\S}^I&=&\left(\frac{\S_{(\psi K_L)\perp, l^-X}}{\kappa_{(\psi K_L)\perp, l^-X}}
+\frac{\S_{(l^+X)\perp, \psi K_S}}{\kappa_{(l^+X)\perp, \psi K_S}}\right)
-\frac{\Delta N^{I}_R}{2}\left(\frac{\S_{(\psi K_L)\perp, l^-X}}{\kappa_{(\psi K_L)\perp, l^-X}}
-\frac{\S_{(l^+X)\perp, \psi K_S}}{\kappa_{(l^+X)\perp, \psi K_S}}
\right) \nn \\ 
&\simeq&
2G(z_K^I-\Delta \lambda_{\mathrm{wst}}^I)+2(S^2-1)z^I+
2GS\hat{\lambda}_l^R+S\Delta N^{I}_R\nn\\
&=&
2[G(z_K^I-\Delta\lambda_{\mathrm{wst}}^I)-z^I+SR_M
+S\hat{\lambda}_{\mathrm{wst}}^R-SC_\xi^l-S\xi_l^R],
\label{A11}\\
\hat{\C}^I&\simeq&\frac{\C_{(\psi K_L)\perp, l^-X}}{\kappa_{(\psi K_L)\perp, l^-X}}
+\frac{\C_{(l^+X)\perp, \psi K_S}}{\kappa_{(l^+X)\perp, \psi K_S}}
=
2[z_K^R-\Delta\lambda_{\mathrm{wst}}^R-Gz^R-S\hat{\lambda_l^I}].
\label{A12}
\eea
\section{The relation among coefficients of the asymmetries for processes 
(I)-(IV).}
\label{F}
In this appendix, we show the relation among the coefficients for different processes (I-IV).
First, we note the coefficients of the process II(IV) are obtained by changing the sign of the mixing parameter $q_K$ and $z_K$ of I(III).
The change of the sign of $q_K$ leads to the
change of the sign for $S, G$ and $\lambda_{{\rm wst}}$.
Next, we show a simple rule which enables one to obtain the coefficients for table \ref{Tab4},
with the coefficients of table \ref{Tab6}.  
For this purpose we do not substitute $\pm 1$ for $C_{l^{\pm}}$ respectively and
write the coefficients of asymmetry for process IV,
\bea
   R^{IV}_T&=&-Sz^I+\frac{1}{2}(C_{l^+}-C_{l^-})R_M-\xi_l^R-C_{\xi}^l+\hat{\lambda}_{\mathrm{wst}}^R-G\hat{\lambda}_l^R\label{F1},\\
   C^{IV}_T&=&\frac{1}{2}(C_{l^-}-C_{l^+})C+Sz^I+\frac{1}{2}(C_{l^-}\theta^R_{K_L}-C_{l^+}\theta^R_{K_S})-S\Delta\lambda_l^I,\\
   S^{IV}_T&=&\frac{1}{2}(C_{l^-}-C_{l^+})S+SGz^R+\frac{G}{2}(C_{l^+}\theta_{K_S}^I-C_{l^-}\theta_{K_L}^I)\\
   &&-GS(C_{l^+}\mathrm{Re}[\lambda_{l^+}]+C_{l^-}\mathrm{Re}[\lambda^{-1}_{l^-}]),\\  
   B^{IV}_T&=&S[-Gz_K^I+z^I+\frac{C_{l^-}-C_{l^+}}{2}SR_M+S\xi_l^R]+S^2C_{\xi}^l-S^2\hat{\lambda}_{\mathrm{wst}}^R+SG\Delta\lambda_{\mathrm{wst}}^I,\\
   D^{IV}_T&=&S[z_K^R-Gz^R]-S\Delta\lambda_{\mathrm{wst}}^R+\frac{C_{l^-}-C_{l^+}}{2}S^2\hat{\lambda}_l^I,\\
   E^{IV}_T&=&\frac{C_{l^-}-C_{l^+}}{2}GS.\label{F7}
\eea
When $l^+$ and $l^-$ in eq. (\ref{A3}) are exchanged, 
the sign of $C_{\xi}^l$ and $\xi_l^R$ is reversed.
According to eqs. (\ref{Lep9}-\ref{Lep10}), the sign of $\hat{\lambda}_l^I$ and $\Delta\lambda_l^R$ 
also changes.
Additionally, one needs to interchange $C_{l^+}$ and $C_{l^-}$
in eqs. (\ref{F1}-\ref{F7})
and one can obtain the coefficients of asymmetry for process II.
\section{
Calculation of equivalence conditions
}
In this appendix, we give the derivation of eqs. (\ref{J2}-\ref{J4}).
The expression of the final state of signal side
in figure (\ref{Fig1}) is,
\bea
 \ket{B_{(\rightarrow \psi K_S)\perp}}=
 N_{(\rightarrow \psi K_S)\perp}
 (\bar A_{\psi K_S}\ket{B^0}-A_{\psi K_S}\ket{\bar B^0}),\label{Bket}
\eea
since the state is orthogonal to $\bra{\psi K_S}$.
The state orthogonal to $\ket{\psi K_L}$ is
\bea
\bra{B_{(\psi K_L\rightarrow)\perp}}=N_{(\psi K_L \rightarrow)\perp}(\bar A^{\mathrm{ID}}_{\psi K_L}\bra{B^0}-A^{\mathrm{ID}}_{\psi K_L}\bra{\bar B^0}).\label{Bbra}
\eea
Similarly, one can write down,
\bea
\bra{B_{(\psi K_S\rightarrow)\perp}}&=&N_{(\psi K_S\rightarrow)\perp}
(\bar{A}_{\psi K_S}^\mathrm{ID}\bra{B^0}-A_{\psi K_S}^\mathrm{ID}\bra{\bar{B^0}}),
\label{EE7}\\
\ket{B_{(\rightarrow \psi K_L)\perp}}&=&N_{(\rightarrow \psi K_L)\perp}
(\bar{A}_{\psi K_L}\ket{B^0}-A_{\psi K_L}\ket{\bar{B^0}})
\label{EE8}.
\eea
Calculating the inner product of eqs. (\ref{Bket}) and (\ref{Bbra}), we obtain
\bea
\braket{B_{(\psi K_L\rightarrow)\perp}|B_{(\rightarrow \psi K_S)\perp}}&=&N_{(\rightarrow \psi K_S)\perp}N_{(\psi K_L \rightarrow)\perp}
  (\bar A_{\psi K_S}\bar A^{\mathrm{ID}}_{\psi K_L}+A_{\psi K_S}A^{\mathrm{ID}}_{\psi K_L})\nn \\
&=&\frac{1}{2}N_{(\rightarrow \psi K_S)\perp}N_{(\psi K_L \rightarrow)\perp}[A_{\psi K^0}A_{\psi K^0}^{\mathrm{ID}}-\bar{A}_{\psi \bar{K^0}}\bar{A}_{\psi \bar{K^0}}^{\mathrm{ID}}\nn\\
&&-\frac{q_K}{p_K}(A_{\psi K^0}A_{\psi \bar K^0}^{\mathrm{ID}}+\bar{A}_{\psi \bar{K^0}}^{\mathrm{ID}}\bar{A}_{\psi K^0})
+\frac{p_K}{q_K}(A_{\psi \bar K^0}A_{\psi K^0}^{\mathrm{ID}}+\bar{A}_{\psi \bar{K^0}}\bar{A}_{\psi K^0}^{\mathrm{ID}})]\nn\\
&=&\frac{N_{(\rightarrow \psi K_S)\perp}N_{(\psi K_L \rightarrow)\perp}}{2}(A_{\psi K^0}A_{\psi K^0}^{\mathrm{ID}}+\bar{A}_{\psi \bar{K^0}}\bar{A}_{\psi \bar{K^0}}^{\mathrm{ID}})
 [\theta_K+\Delta\lambda_{\mathrm{wst}}],\nn\\
\label{innerP}
\eea
where we used eqs. (\ref{D1}-\ref{D8}).
The inner product in eq.(\ref{innerP}) was previously obtained
in \cite{Applebaum:2013wxa}.
In eq.(\ref{innerP}), we compute it with our notation including the normalization constant and have
ignored the second order of small parameters\\
$z_K, \theta_{\psi K^0}, \bar{\theta}_{\psi \bar{K^0}}, \hat \lambda_{\mathrm{wst}}
$ and
$\Delta\lambda_{\mathrm{wst}}$.\\
Next, we show the derivation of the first line of eq. (\ref{J2})
and the second line of eq. (\ref{J4}).
The states are given as,
\bea
\bra{B_{(l^-X\rightarrow )\perp}}&=&N_{(l^-\rightarrow )\perp}(\bar A_{l^-}^{\mathrm{ID}}\bra{B^0}-A_{l^-}^{\mathrm{ID}}\bra{\bar{B^0}})\label{EE5},\\
 \ket{B_{(\rightarrow l^+X)\perp}}&=&N_{(\rightarrow l^+)\perp}(\bar A_{l^+}\ket{B^0}-A_{l^+}\ket{\bar{B^0}})
\label{EE6},\\
\bra{B_{(l^+X\rightarrow )\perp}}&=&N_{(l^+\rightarrow )\perp}(\bar A_{l^+}^{\mathrm{ID}}\bra{B^0}-A_{l^+}^{\mathrm{ID}}\bra{\bar{B^0}})\label{EE9},\\
 \ket{B_{(\rightarrow l^-X)\perp}}&=&N_{(\rightarrow l^-)\perp}(\bar A_{l^-}\ket{B^0}-A_{l^-}\ket{\bar{B^0}})
\label{EE10}.
\eea
Their inner product is,
\bea
 \braket{B_{(l^-X\rightarrow )\perp}|B_{(\rightarrow l^+X)\perp}}&=&N_{(l^-\rightarrow )\perp}N_{(\rightarrow l^+)\perp}
 (\bar A_{l^-}^{\mathrm{ID}}\bar A_{l^+}+A_{l^-}^{\mathrm{ID}}A_{l^+})\nn\\
 &=&2N_{(l^-\rightarrow )\perp}N_{(\rightarrow l^+)\perp}A_{l^+}\bar A_{l^-}^{\mathrm{ID}}\frac{p}{q}\lambda_{l^+}.
\eea
The proportionality to the wrong sign decay amplitude $\lambda_{l^+}$ is derived in \cite{Applebaum:2013wxa}.
\acknowledgments
We would like to thank Dr. T. Shindou for useful comments and suggestion,
and helpful correspondence by A. Efrati, Y. Nir and Y. Soreq.

%


\begin{thebibliography}{99}
\bibitem{Banuls:1999aj} 
  M.~C.~Banuls and J.~Bernabeu,
  CP, T and CPT versus temporal asymmetries for entangled states of the B(d) system,
  Phys.\ Lett.\ B {\bf 464}, 117 (1999)
  [hep-ph/9908353].
\bibitem{Alvarez:2006nk} 
  E.~Alvarez and A.~Szynkman,
  Direct test of time reversal invariance violation in B mesons,
  Mod.\ Phys.\ Lett.\ A {\bf 23}, 2085 (2008)
  [hep-ph/0611370].
\bibitem{Bernabeu:2012ab} 
  J.~Bernabeu, F.~Martinez-Vidal and P.~Villanueva-Perez,
  Time Reversal Violation from the entangled B0-antiB0 system,
  JHEP {\bf 1208}, 064 (2012)
  [arXiv:1203.0171 [hep-ph]].
\bibitem{Lees:2012uka} 
  J.~P.~Lees et al. [BaBar Collaboration],
  Observation of Time Reversal Violation in the $B^0$ Meson System,
  Phys.\ Rev.\ Lett.\  {\bf 109}, 211801 (2012)
  [arXiv:1207.5832 [hep-ex]].
\bibitem{Schubert:2014ska} 
  K.~R.~Schubert,
  T Violation and CPT Tests in Neutral-Meson Systems,
  arXiv:1409.5998 [hep-ex].
\bibitem{Applebaum:2013wxa} 
  E.~Applebaum, A.~Efrati, Y.~Grossman, Y.~Nir and Y.~Soreq,
  Subtleties in the BaBar measurement of time-reversal violation,
  Phys.\ Rev.\ D {\bf 89}, 076011 (2014)
  [arXiv:1312.4164 [hep-ph]].
\bibitem{Grossman:2002bu} 
  Y.~Grossman, A.~L.~Kagan and Z.~Ligeti,
  Can the CP asymmetries in B$\rightarrow$ $\psi$ K(s) and B $\rightarrow$ $\psi$ $K(L)$ differ?,
  Phys.\ Lett.\ B {\bf 538}, 327 (2002)
  [hep-ph/0204212].
\bibitem{Lenz:2011ti} 
  A.~Lenz and U.~Nierste,
  Numerical Updates of Lifetimes and Mixing Parameters of B Mesons,
  arXiv:1102.4274 [hep-ph].
\bibitem{Bigi:2000yz} 
  I.~I.~Y.~Bigi and A.~I.~Sanda,
  CP violation,
  Camb.\ Monogr.\ Part.\ Phys.\ Nucl.\ Phys.\ Cosmol.\  {\bf 9}, 1 (2000).
\bibitem{Li:2006vq} 
  H.~n.~Li and S.~Mishima,
  Penguin pollution in the $B^0 \rightarrow J/\psi K(S)$ decay,
  JHEP {\bf 0703}, 009 (2007)
  [hep-ph/0610120].
\bibitem{Dighe:2001gc} 
  A.~S.~Dighe, T.~Hurth, C.~S.~Kim and T.~Yoshikawa,
  Measurement of the lifetime difference of B(d) mesons: Possible and worthwhile?,
  Nucl.\ Phys.\ B {\bf 624}, 377 (2002)
  [hep-ph/0109088].
\bibitem{Dighe:2001sr}
  A.~Dighe, T.~Hurth, C.~S.~Kim and T.~Yoshikawa,
  The Width difference of B(d) mesons,
  PoS HEP {\bf 2001}, 096 (2001)
  [hep-ph/0112067].
\bibitem{Sachs:1963zz} 
  R.~G.~Sachs,
  Methods for Testing the CPT Theorem,
  Phys.\ Rev.\  {\bf 129}, 2280 (1963).
\bibitem{Enz:1965tr} 
  C.~P.~Enz and R.~R.~Lewis,
  On the phenomenological description of CP violation for K mesons and its consequences,
  Helv.\ Phys.\ Acta {\bf 38}, 860 (1965).
\bibitem{AlvarezGaume:1998yr} 
  L.~Alvarez-Gaume, C.~Kounnas, S.~Lola and P.~Pavlopoulos,
  Violation of time reversal invariance and CPLEAR measurements,
  Phys.\ Lett.\ B {\bf 458}, 347 (1999)
  [hep-ph/9812326].
\bibitem{Beuthe:1997fu} 
  M.~Beuthe, G.~Lopez Castro and J.~Pestieau,
  Field theory approach to K0 - anti-K0 and B0 - anti-B0 systems,
  Int.\ J.\ Mod.\ Phys.\ A {\bf 13}, 3587 (1998)
  [hep-ph/9707369].
\bibitem{Silva:2000db} 
  J.~P.~Silva,
  On the use of the reciprocal basis in neutral meson mixing,
  Phys.\ Rev.\ D {\bf 62}, 116008 (2000)
  [hep-ph/0007075].
\end{thebibliography}
\end{document}